\documentclass[usenatbib]{mnras}  

\usepackage{newtxtext,newtxmath}

\usepackage[T1]{fontenc}

\DeclareRobustCommand{\VAN}[3]{#2}
\let\VANthebibliography\thebibliography
\def\thebibliography{\DeclareRobustCommand{\VAN}[3]{##3}\VANthebibliography}

\usepackage{subcaption} 
\usepackage{rotating}   
\usepackage{graphicx}	
\usepackage{amsmath}	
\usepackage{amssymb}	
\usepackage{lscape}     
\usepackage{cases}      
\usepackage{multirow}   
\usepackage{tikz}       
\usepackage{overpic}    
\usepackage{anyfontsize}
\usepackage[normalem]{ulem}
\newcommand{\KI}{\texttt{ME}}
\newcommand{\KR}{\texttt{MT}}
\newcommand{\DC}{\texttt{DC}}

\title[Impact of subgrid physics on the CGM]{ARCHITECTS I: Impact of subgrid physics on the simulated properties of the circumgalactic medium}

\author[M. Rey et al.]{Maxime Rey$^{1,2}$\thanks{E-mail: maximerey.astro@gmail.com},
Jérémy Blaizot$^{2}$,
Taysun Kimm$^{1}$,
Joakim Rosdahl$^{2}$,
Léo Michel-Dansac$^{3}$
\\
$^{1}$Center for Galaxy Evolution Research and Department of Astronomy, Yonsei University, 50 Yonsei-ro, Seodaemun-gu, Seoul 03722, Republic of Korea \\
$^{2}$Centre de Recherche Astrophysique de Lyon UMR5574, ENS de Lyon, Univ. Lyon, CNRS, 9 av. Charles André, F-69230 Saint-Genis-Laval, France \\
$^{3}$Aix Marseille Univ., CNRS, CNES, LAM, Marseille, France\\
}

\date{Accepted XXX. Received YYY; in original form ZZZ}

\pubyear{2022}

\begin{document}
\label{firstpage}
\pagerange{\pageref{firstpage}--\pageref{lastpage}}
\maketitle

\begin{abstract}
Galaxy evolution is shaped by star formation and stellar feedback at scales unresolved by current high-resolution cosmological simulations. Precise subgrid models are thus necessary, and different approaches have been developed. However, they are degenerate and often primarily calibrated to reproduce stellar masses from observations.
To explore these degeneracies, we perform three cosmological zoom-in radiation-hydrodynamics simulations of the same galaxy within a $5\times10^{11}\rm\ M_\odot$ dark matter halo at $z\sim1$, each with a different subgrid model: mechanical feedback, a combination of mechanical feedback and thermal feedback, and delayed cooling. We calibrate the simulations to match in stellar mass, isolating the effect of the models on the circumgalactic medium (CGM).
Our findings demonstrate that despite producing galaxies with comparable stellar masses, the three models lead to distinct feedback modes, resulting in notable variations in the CGM properties. The delayed cooling run is dominated by ejective feedback and exhibits high burstiness, whereas mechanical and the hybrid models primarily feature preventive feedback, respectively acting at the galaxy and halo scales. Delayed cooling reduces the baryon mass to half the universal baryon fraction while mechanical feedback retains most baryons, with the hybrid model standing in between. Delayed cooling also ejects significantly more metals into the CGM than both other models. While for delayed cooling and mechanical feedback metals are almost evenly distributed in the CGM, they are concentrated around satellites in the hybrid model. These discrepancies emphasize the need to design an appropriate subgrid model to understand how stellar feedback regulates galaxy growth.

\end{abstract}



\section{Introduction}
By allowing the simultaneous modelling of nonlinear processes across vast ranges of spatial and temporal scales, numerical simulations have become an essential tool to study the complex processes that govern galaxy formation and evolution. In recent years, significant progress has been made in creating simulations that describe large, representative volumes of the Universe, yielding results consistent with many observations \citep[e.g.][]{Dubois2014, Schaye2015, Nelson2019, Dave2019}. 
However, these simulations rely on macroscopic subgrid models that describe the behaviour of the interstellar medium (ISM) as a whole, prompting the development of a new generation of simulations, aimed at a deeper understanding of the fundamental physics shaping galaxy evolution \citep{Hopkins2014, Hopkins2018, Hopkins2023, Wang2015, Grand2017, Agertz2020}. These simulations focus on zooming in on specific halos within a cosmological box, preserving large-scale structures while achieving higher resolution in the galaxy.
While these simulations reach much higher resolutions, they require the development of new models to describe still unresolved processes like star formation and feedback. Accurately capturing supernovae and mitigating the overcooling problem requires a higher resolution than typically achieved in modern zoom simulations: $\approx 4\rm\ pc$ \citep{Kim2015, Simpson2015} in Eulerian simulations. If these conditions are not met, most of the supernova’s energy is lost before a shockwave can form \citep{Katz1992, Cho2008, Ceverino2009, Chaikin2022}. One method used to overcome this limitation is delayed cooling \citep{Stinson2006, Governato2010, Agertz2011, Teyssier2013}, where cooling is temporarily disabled to prevent spurious energy loss immediately after a supernova. More recently, \citet{Kimm2014, Kimm2015} introduced mechanical feedback \citep[see also][]{Hopkins2014}, where energy derived from high-resolution simulations is injected as momentum. Some approaches, such as that of \citet{Kretschmer2020}, combine thermal energy and mechanical feedback.

The impact of subgrid models is often tested in either high-resolution idealised galaxy simulations \citep{Hopkins2013, Rosdahl2017, Smith2018, Smith2021b, Andersson2023} or in zoom-in cosmological simulations \citep{Kimm2014, Nunez-Castineyra2021, Azartash-Namin2024, Kang2025}. Although different models can yield similar galaxy properties, they can still produce distinct feedback mechanisms such as ejective, where gas is expelled from the galaxy, preventive feedback, which prevents gas from falling onto the galaxy or galactic fountains where gas that has been ejected eventually falls back onto the galaxy \citep{Christensen2016, Mitchell2022}. Discriminating between these models and refining them remains a key challenge for new-generation simulations.
Most of these comparative studies focus solely on the stellar or gaseous properties of the galactic disk, often overlooking the crucial role of the circumgalactic medium (CGM). This highly multiphase, multiscale environment \citep{Tumlinson2017, Faucher-Giguere2023, Sameer2024, Peroux2024} surrounds galaxies and this is where inflows from cosmological accretion interact with feedback-driven outflows \citep{Crain2013, Hafen2019, Afruni2021, Decataldo2024, Saeedzadeh2023}. Understanding how these flows are regulated is essential to unveil the processes that govern star formation beyond ISM scales. Given its critical role, many studies investigate the properties of the CGM \citep{Mitchell2018, Afruni2021, Afruni2023b, Decataldo2024}, and observations of the CGM are becoming increasingly detailed \citep{Fumagalli2024, Chen2024}, offering complementary constraints that could help differentiate between models.\\
Simulations studying the CGM must incorporate a cosmological context, as seen in \citet{Davies2020, Oppenheimer2020, Medlock2025}, where CGM properties were compared across the \textsc{EAGLE} \citep{Crain2015, Schaye2015}, \textsc{SIMBA} \citep{Dave2019}, and \textsc{IllustrisTNG} \citep{Pillepich2019, Nelson2019b} cosmological simulations.
However, full cosmological simulations typically lack ISM resolution, and zoom-in simulations address this by focusing on a single halo at high resolution, though at the cost of the loss of large-number statistics. \citet{Kelly2022} compared the CGM of two simulations suites using this approach, namely the \textsc{AURIGA} \citep{Grand2017} and \textsc{APOSTLE} simulations \citep{Sawala2016, Fattahi2016}. Nevertheless, the minimum softening length in these simulations is only about $300\rm\ pc$. To the best of our knowledge, this represents the highest resolution comparison of the impact of subgrid models on the CGM to date. Moreover, in CGM comparisons from both full cosmological simulations \citep{Davies2020, Oppenheimer2020, Medlock2025} and zoom-in simulations \citep{Kelly2022}, not only do subgrid physics change, but also the entire simulation setup, such as the code, the settings (e.g. the resolution), and initial conditions, making it difficult to isolate the specific impact of subgrid models on the CGM. 
In this paper, we take a similar approach but focus on the impact of star formation and feedback models on the CGM with much higher resolution. We do not aim to identify the best subgrid model, but rather to emphasize the critical role these models play in shaping the properties of the CGM, even when they produce similar stellar masses.

To this end, we introduce the \textsc{ARCHITECT} suite of simulations, designed to Assess the Robustness of Computational Hydrodynamics and ISM-scale Theories by Exploiting Circumgalactic Tracers. We run three high-resolution radiation-hydrodynamics simulations, based on identical initial conditions, differing only in the subgrid models employed. Each of the three simulations uses a different set of star formation and supernova feedback models often used in the literature, which we refer to as \KI\ \citep[mechanical feedback,][]{Kimm2017}, \KR\ \citep[mechanical + thermal,][]{Kretschmer2020}, and \DC\ \citep[delayed cooling,][]{Teyssier2013}, as described in section \ref{sec:methods}.
Since CGM properties are expected to depend on galaxy mass, as indicated by both observational \citep{Bordoloi2018, Dutta2021} and theoretical studies \citep{Nelson2020, Anand2021, Cook2024}, we calibrate the stellar mass of the three galaxies to isolate the impact of feedback. With similar galaxies in place, we present their properties and compare the effects of the three subgrid models on the CGM in section \ref{sec:glob_prop}, before concluding on our findings in section \ref{sec:conclusions}.

\section{Numerical methods} \label{sec:methods}
    In this section, we briefly describe the code we use to run our simulations and how we generate the initial conditions. We also list the physics included in the simulations and detail how the subgrid models work, focusing on the differences between the three approaches.

    \subsection{Numerical setup}
        The simulation setup and physics included closely follow those used in \textsc{SPHINX} \citep{Rosdahl2018}. We perform our simulations with the hydrodynamical code RAMSES \citep{Teyssier2002, Teyssier2010}, relying on an Adaptive Mesh Refinement strategy and based on the Euler conservation equations in the presence of self-gravity and cooling. Non-equilibrium cooling is included alongside ionising radiative transfer \citep{Rosdahl2013, Rosdahl2015b}.
        Dark matter and stars are modelled as collisionless particles interacting gravitationally through a particle-mesh solver and cloud-in-cell interpolation \citep{Guillet2011}. Their minimum masses are respectively $m_{\rm DM} = 3.49 \times 10^5\ \rm M_\odot$ and $m_\star = 3.2 \times 10^3\ \rm M_\odot$.

        We use MUSIC \citep{Hahn2011, Hahn2013} to generate cosmological initial conditions for the simulations presented here, assuming cosmological parameters $\Omega_\Lambda = 0.6825$, $\Omega_\mathrm{m} = 0.3175$, and $\Omega_\mathrm{b} = 0.049$ \citep{PlanckCollaboration2014}. 
        Our zoom-in simulations describe a halo of $M_{\rm h} = 5.33 \times 10^{11}\ \rm M_\odot$ at $z=0$, which was selected in a $30\rm\ cMpc\,h^{-1}$ box to be relatively isolated, with no massive sub-structure and no neighbour more massive than $0.2\times M_{\rm h}$ within $3\ R_{\rm vir}$, with $R_{\rm vir}$ the virial radius. The zoom-in region extends beyond $3\ R_{\rm vir}$ at all redshifts down to $z=1$. The resolution of our simulation ranges from level 7 ($128^3$ particles/cells at the coarse level) to level 20 (
        $8192^3$ effective resolution, $\Delta x\approx 40\rm\ pc$) in the zoom region.
        
        \begin{figure}
            \begin{subfigure}[h]{\columnwidth}
                \hspace{0.13\columnwidth}
                \includegraphics[width=0.845\columnwidth]{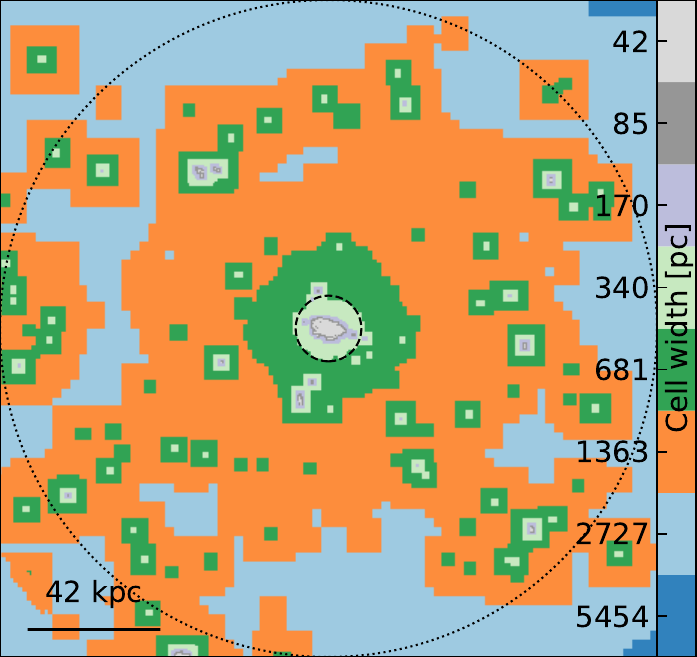}
            \end{subfigure}
            \begin{subfigure}[h]{\columnwidth}
        	    \includegraphics[width=\columnwidth]{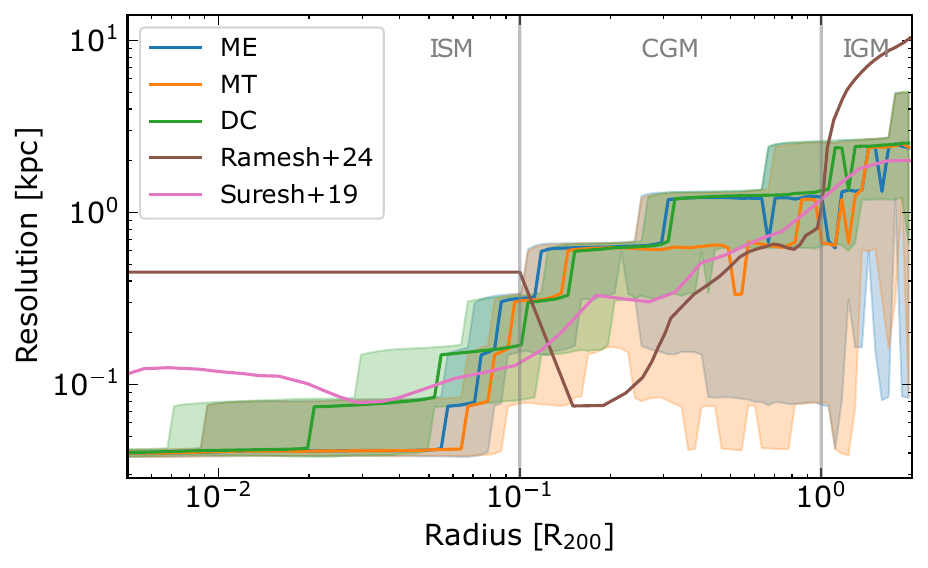} 
            \end{subfigure}
            \caption{Top: Projection of the cells width for one of the \textsc{ARCHITECT} simulations (\KI) at redshift $z=1$ in parsec. We trace two circles in dashed and dotted lines, with a respective radius of $0.1 \,R_{\rm 200}$ and $R_{\rm 200}$. Bottom: cell width as a function of radius in units of $R_{\rm 200}$ for the three \textsc{ARCHITECT} simulations (\KI, \KR, and \DC, presented hereafter). The solid line shows the median resolution over $1\rm\ Gyr$ (from $z=1.3$ to $z=1$) and the shaded area corresponds to the 15.9 and 84.1 percentiles. We compare our simulation's resolution to \citet{Suresh2019} and \citet{Ramesh2024c}.}
            \label{fig:radial_res}
        \end{figure}
        To illustrate how our simulations compare to other simulations in terms of resolution, we show in Fig.~\ref{fig:radial_res} a projection of the cell width of one of our runs (\KI), as well as the cell width profile of the three \textsc{ARCHITECT} simulations as a function of radius. We normalise the x-axis of the bottom panel by $R_{\rm 200}$, the radius within which the mean density of the medium is 200 times the critical density of the Universe. We compare our resolution to \citet{Suresh2019} (the \textsc{L11\_12} simulation) and \citet{Ramesh2024c} (the \textsc{GIBLE} simulation), two zoom simulations which aimed at increasing the CGM resolution to resolve cold gas structures. Beyond $ R_{\rm 200}$ and down to $\approx 0.5\,R_{\rm 200}$, our simulations are close to those of \textsc{L11\_12}, although \KI\ and \DC\ have somewhat worse resolution than \KR. In the inner CGM ($\lesssim 0.5\,R_{\rm 200}$), our simulation has a cell width twice lower than \textsc{L11\_12}, and two to eight times lower than \textsc{GIBLE}, the highest difference being close to the galaxy, at $0.2\,R_{\rm 200}$. At lower radii, while the resolution decreases sharply in \textsc{GIBLE} to reach $450\rm\ pc$ in the ISM, our simulations exhibit a sharp increase in resolution down to a roughly ten times higher resolution than \textsc{GIBLE}, at $40\rm\ pc$. \textsc{L11\_12} exhibits similar ISM resolution to our simulation, albeit a bit lower, around $100\rm\ pc$.

        \vspace{0.3 cm}
        We use spectral energy distributions from the Binary Population and Spectral Synthesis code \citep[v2.2.1, ][]{Stanway2018} and trace the ionising radiation in the simulations using three photon groups defined by the ionisation energies of H$\,\textsc{i}$, He$\,\textsc{i}$ and He$\,\textsc{i}$: [13.6, 24.59), [24.59, 54.42), [54.42, $\infty$). We do not account for radiation below the hydrogen ionisation energy. The interactions simulated for these groups are photoionisation, cooling and heating, and momentum transfer \citep{Rosdahl2015b}. 
        Radiation flows through the grid and thus constrains the Courant condition. We hence use the reduced speed of light approximation with $\mathrm{c_{sim}} = 0.005\rm\,c$ \citep{Gnedin2001, Rosdahl2013} to limit the computational cost.
        We turn on a uniform ionisation UV field at $z=8.5$, modelling emission from unresolved surrounding galaxies, following \citet{Faucher-Giguere2009}, and take into account self-shielding with an exponential damping factor above densities of $10^{-2}\rm\ cm^{-3}$ \citep{Schaye2001, Rosdahl2012b}.

        We trace the ionisation states of atomic hydrogen and helium, and track metals through the single variable $Z$, describing the metal mass fraction. Non-equilibrium cooling is computed for hydrogen and helium via photoionisation, collisional ionisation and excitation, recombination, bremsstrahlung, homogeneous Compton cooling/heating off the cosmic microwave background, and dielectronic recombination. Above $10^4\rm\ K$, metal cooling is modelled following \textsc{CLOUDY} tables (\citealp{Ferland1998}, version 6.02) assuming photoionisation equilibrium with the UV background from \citet{Haardt1996}. Below $10^4\rm\ K$, fine structure metal cooling rates from \citet{Rosen1995} are used down to $10\rm\ K$.

        \vspace{0.3 cm}
        In the zoom-in region, two criteria can trigger refinement in the AMR grid. The first criterion follows density and increases the resolution when a cell's mass (DM and baryonic) is greater than a selected threshold. For the cell to be refined, the condition is $m_{\rm DM, cell} + \frac{\Omega_\mathrm{m}}{\Omega_\mathrm{b}} m_{\rm b, cell} \geq 8 m_{\rm DM}$ where $m_{\rm DM, cell}$ is the dark matter mass in the cell, $m_{\rm b, cell}$ the baryonic mass in a cell (gas + stellar mass) and $m_{\rm DM}$ the mass of a single dark matter particle.
        The second refinement criterion aims to keep the local thermal Jeans length $\lambda_\mathrm{J}$ resolved by at least 4 cells, down to the resolution limit. Thus, a cell is refined whenever 
        \begin{equation} 
        4\Delta x \geq \lambda_\mathrm{J} = \sqrt{\frac{\pi c_\mathrm{s}^2}{\rho G}}, \label{eq:Jeans_length_refinement}
        \end{equation} 
        with $\rho$ the gas density, $G$ the gravitational constant, and $c_\mathrm{s}$ the sound speed.

        \vspace{0.3 cm}
        Feedback from massive stars in the form of photoionisation, heating and radiation pressure from young stars is accounted for in our simulations via the radiation-hydrodynamics solver \citep{Rosdahl2013, Rosdahl2015b}. We nevertheless need two subgrid models to describe the formation of stars and the effect of explosions of type II supernovae. In the present paper, we test and compare 3 models for this, which we describe in Secs.~\ref{subsec:KIsim} to \ref{subsec:DCsim}.

    \subsection{\KI\ simulation}  \label{subsec:KIsim}
        \KI\ uses a Schmidt-based, thermo-turbulent subgrid model for star formation \citep{Kimm2017}, and mechanical supernova feedback by \citet{Kimm2015}. This model is used in many simulations such as \textsc{SPHINX} \citep{Rosdahl2018, Rosdahl2022}, \textsc{SPHINX-MHD} \citep{Katz2021}, NewHorizon \citep{Dubois2014}, \textsc{OBELISK} \citep{Trebitsch2021} or the \textsc{NewHorizon2} model \citep{Yi2024}. We first summarise how the multi-free-fall model for star formation is applied in \KI. Then, we describe the radiative feedback used in all our simulations alongside the modelling of supernovae feedback in \KI.

        \subsubsection*{Star formation}
            Star formation can be broadly described in three steps. (1) determining whether a cell is star-forming, (2) calculating the local SFR, and (3) implementing the SFR over a timestep dt using finite-mass star particles.

            In \KI, we allow star formation if a cell has a gas hydrogen number density higher than $10\rm\ cm^{-3}$, is a local maximum in density compared to its six neighbouring cells, is in a converging flow (${\Vec{\nabla} \cdot (\rho \Vec{v})} \leq 0$), and has a size $\Delta x \geq \lambda_{\rm J, turb}/4$, where $\lambda_{\rm J, turb}$ is the turbulent Jeans length:
            \begin{equation} \label{eq:lambda_jeans_SF}
                \lambda_{\rm J, turb}(\rho)=\frac{\pi \sigma_{\rm gas}^2 \pm \sqrt{36\pi c_\mathrm{s}^{2} G \Delta x^{2}\rho + \pi^{2}\sigma_{\rm gas}^4}}{6 G \Delta x \rho}.
            \end{equation}
            The three-dimensional gas velocity dispersion $\sigma_\mathrm{gas}$ in Eq.~\ref{eq:lambda_jeans_SF} is computed locally as the norm of the gradient of the velocity field from the neighbouring cells. 

            Once a cell is flagged for star formation, it will form stars at a rate given by
            \begin{equation} \label{eq:gas2stars}
                \dot{\rho}_{\star} = \rho \frac{\epsilon_\mathrm{ff}}{t_\mathrm{ff}},
            \end{equation} 
            with $t_\mathrm{ff} = [{3 \pi}/{(32 G \rho)}]^{1 / 2}$ the free-fall time. The star formation efficiency per free-fall time ($\epsilon_\mathrm{ff}$) is modelled following the multi-free-fall approach laid out by \citet{Hennebelle2011} and \citet{Federrath2012}. The main assumption is that star formation occurs in a supersonic turbulent medium so that the unresolved density distribution in the cell follows a log-normal function. 
            Then, this distribution can be integrated over a critical density $s_\mathrm{crit}$ to define the star formation efficiency 
            \begin{equation} \label{eq:SFR_ff}
                \epsilon_\mathrm{ff} = \epsilon_\mathrm{loc} \exp \left(\frac{3}{8} \sigma_\mathrm{s}^{2}\right)\left[1+\operatorname{erf}\left(\frac{\sigma_\mathrm{s}^{2}-s_\mathrm{crit}}{\sqrt{2 \sigma_\mathrm{s}^{2}}}\right)\right].
            \end{equation}
            Here, $\operatorname{erf}$ is the Gauss error function and $\sigma_\mathrm{s}$ is the standard deviation of the density distribution. We write ${\sigma_\mathrm{s}^{2}=\ln \left(1+b^{2} \mathcal{M}^{2}\right)}$, where $\mathcal{M}$ is the Mach number, and $b=0.4$ the turbulent forcing parameter that describes the ratio of solenoidal to compressive modes. The Mach number $\mathcal{M} = {\sigma_{\rm gas}}/{c_\mathrm{s}}$ depends on the sound speed $c_\mathrm{s}$ and the velocity dispersion $\sigma_{\rm gas}$. The model from \citet{Padoan2011}, which we use in \KI, uses
            \begin{equation} \label{eq:scrit_Ki}
                s_{\rm crit}=\ln \left[0.62 \alpha_{\rm vir} \mathcal{M}^{2}\right],
            \end{equation} 
            where
            \begin{equation} \label{eq:alpha_vir}
                \alpha_{\rm vir} = \frac{5}{\pi \rho G}\frac{\sigma_{\rm gas}^2+c_\mathrm{s}^2}{\Delta x^2}
            \end{equation}
            is the virial parameter describing the gravitational stability of the cloud. The \textit{local} star formation efficiency $\epsilon_\mathrm{loc}=0.1425$  is a constant used to account for unresolved physics such as protostellar jets removing gas from the star-forming region and the estimated error on the timescale for the gas to become unstable \citep{Krumholz2005}. When the Mach number gets large, turbulent compression gets more efficient and the global star formation efficiency $\epsilon_\mathrm{ff}$ increases. Conversely, as the virial parameter gets larger, the cloud gets more gravitationally unstable and $\epsilon_\mathrm{ff}$ decreases \citep[see Fig.~1 in][]{Federrath2012}.

            Finally, the mass of stars formed is $N$ times $m_\star$, where $N$ is an integer number drawn from a Poisson distribution of mean $SFR\times \Delta t$ \citep{Rasera2006}. The stellar particle created represents a simple stellar population defined through its mass, metallicity, age, position, and velocity. We set the minimum mass of the stellar particles in our simulation to $\approx 3200\rm\ M_\odot$. 

        \subsubsection*{Feedback}
            We assume that all type-II supernovae explode 10 Myr after a stellar particle is spawned with a rate of 2 supernovae per $100\rm\ M_\odot$ of stellar mass formed.
            We artificially boost the supernova rate by a factor of two to better reproduce the stellar mass to halo mass relation and match the stellar mass of the other models. Boosting feedback by e.g. increasing the supernova frequency \citep{Rosdahl2018}, decreasing the minimum mass of stars undergoing supernovae \citep{Dubois2021} or increasing the supernovae efficiency \citep{Li2018, Yi2024} is common practice in supernovae models as galaxies otherwise tend to be too massive. This calibration can be seen as compensation for unaccounted physics such as cosmic rays pressure, proto-stellar jets or to model the unresolved clustering of supernovae which can increase the momentum transfer to the ISM \citep{Kim2015, Keller2014}.


            The method used in \citet{Kimm2015} determines in which phase the supernova explosion is when reaching the neighbour cells, depending on the local gas conditions, and injects momentum and energy accordingly. 
            In the snowplough limit where much gas has been swept up ($\sim 100$ times the ejecta), the model injects \citep[and references therein]{Kimm2017}
            \begin{equation}
                \begin{split}
                    p_\mathrm{rad} &= \left[2.5 \times 10^{5} N_\mathrm{SN}^{{16}/{17}} n_\mathrm{H}^{-{2}/{17}} Z'^{-0.14} \right]
                    \mathrm{e}^{-{\Delta x}/{R_\mathrm{st}}}\\
                    &+ \left[5 \times 10^{5} N_\mathrm{SN}^{{16}/{17}} Z'^{-0.14} \right]
                    (1-\mathrm{e}^{-{\Delta x}/{R_\mathrm{st}}}),
                \end{split}
            \end{equation}
            where $N_\mathrm{SN}$ is the number of supernovae at a given timestep, $n_\mathrm{H}$ the density of the neighbouring cell, $Z'=\max(Z,0.01)$, with $Z$ the metallicity in solar units, $\Delta x$ the size of the host cell, and $R_\mathrm{st}$ the Strömgren radius. If the Strömgren radius is not sufficiently resolved, the impact of photoionization is likely to be underestimated. In this case, the second term of $p_\mathrm{rad}$, fitted from \citet{Geen2015}, will dominate the equation. Conversely, if the Strömgren radius is well-resolved, the first term based on \citet{Thornton1998} will prevail.
            %
            In the other limit, where little gas has been swept up, the Sedov-Taylor phase (energy conserving) is well resolved and momentum is injected as
            \begin{equation}
                p_\mathrm{ad, \KI} = B_\mathrm{ph} \sqrt{2 \chi m_\mathrm{ej} f_\mathrm{e} E_\mathrm{SN}},
            \end{equation}
            where $f_\mathrm{e}$ is a coefficient ensuring a smooth transition between the two propagation phases, $B_\mathrm{ph}$ is a boost factor accounting for unresolved photoionisation \citep{Geen2015}, $\chi$ is the ratio of the mass swept up by the supernova to the mass ejected in each cell, $m_\mathrm{ej}$ is the total mass of the ejecta, and $E_\mathrm{SN}$ is the energy released by a single supernova.
            %
            Along with energy and momentum, SN explosions release metals, and we model the mass fraction of metal in the ejecta with a yield of 7.5\%, as $Z_\mathrm{ej} = Z_\mathrm{star}+0.075(1-Z_\mathrm{star})$.
        
    \subsection{\KR\ simulation}
        We now detail the subgrid models used in the \KR\ simulation, which come from \citet{Kretschmer2020}. This model is used for example in the \textsc{MIGA} simulations \citep{Kretschmer2022}, the EMP-Pathfinder suite \citep{Reina-Campos2022} and the \textsc{SPICE} suite of simulations \citep{Bhagwat2024}.

        \subsubsection*{Star formation}
            Here, the model for star formation is also an implementation of the multi-free-fall model, with a number of subtle differences compared to the model from \citet{Kimm2017}. One major difference is the definition of $\sigma_{\rm gas}$, which corresponds to the one-dimensional turbulence.
            \citet{Kretschmer2020} compute turbulence from the velocity and density fields, relying on numerical diffusion to provide an implicit subgrid model \citep{Schmidt2006, Semenov2016}.
            Furthermore, this models sets slightly different parameters values in Eq.~\ref{eq:SFR_ff} (see Table~\ref{tab:models_diff}), and relies on a critical density given by \citep{Krumholz2005}
            \begin{equation} \label{eq:scrit_Kr}
                s_\mathrm{crit}=\ln \left[\alpha_\mathrm{vir}\left(1+\frac{2\mathcal{M}^4}{1+\mathcal{M}^2}\right)\right].
            \end{equation}
            While the role of $\alpha_\mathrm{vir}$ in $s_\mathrm{crit}$ is the same in both \KR\ and in \KI, the Mach number has a distinct impact in the two models. In particular, for large $\mathcal{M}$, $s_\mathrm{crit}$ in \KR\ approaches the form used in \KI, as ${s_\mathrm{crit} \underset{\scriptscriptstyle\mathcal{M}\gg1}{\longrightarrow} \ln \left[2\alpha_\mathrm{vir}\mathcal{M}^2\right]}$, albeit higher by 1.17.  
            Consequently, \KR\ limits star formation to denser media than \KI, leading to a seemingly smaller star formation efficiency. However, this is counterbalanced by a higher local star formation efficiency $\epsilon_\mathrm{loc}=0.5$, as shown in Table~\ref{tab:models_diff}.


        \subsubsection*{Feedback} 
            The original model of \citet{Kretschmer2020} includes a subgrid description for radiative feedback. As we already include RT, we turn this component of the subgrid models off to avoid double-counting.
            Individual supernova explosions are modelled as taking place for each stellar particle at ages ranging from $3\rm\ Myr$ to $20\rm\ Myr$, with a uniform distribution in time.
            The location at which the supernovae explode is chosen randomly over the eight cells comprised in the oct where the stellar particle is located. At each supernova explosion, $10^{51}\rm\ erg$ is injected as thermal energy into the host cell.
            \citet{Kretschmer2020} also compute a local cooling scale  \citep{Martizzi2015}
            \begin{equation}
                R_\mathrm{cool} = 43.6 \,n_\mathrm{H}^{-0.42} \,Z'^{-0.05}\rm\ pc, 
            \end{equation}
            which they compare to the cell size to decide when unresolved cooling losses are important. When $R_{\rm cool} > 4 \Delta x$, \KR\ also inject the momentum expected at the radiative phase as
            \begin{equation}
                p_{\mathrm{rad, \KR}} = 2.66 \times 10^{5} \eta_R N_{\mathrm{SN}} Z'^{-0.114} n_\mathrm{H}^{-0.19}\rm\ M_\odot\,km\,s^{-1}.
            \end{equation}
            $\eta_R$ is a factor dependent on how resolved the cooling radius is. As done for the previous model, we enhance the effect of supernova feedback, this time by boosting the supernova rate by a factor of four compared to the fiducial value from \citet{Kretschmer2020}, resulting in 8 supernovae explosions per $100\rm\ M_\odot$.
            A further difference with \KI\ is that this momentum is not simply injected as momentum in the surrounding cells but also treated as a source term at the solver level, following \citet{Agertz2013}. 
            The ejected mass is obtained by taking the product of the number of supernovae and their average mass and has a metal yield of 10\%.

            \begin{table*}
                \centering
                \caption{
                Summary of the main differences between the subgrid models used in \KI, \KR, and \DC. The three first rows describe the differences in the star formation subgrid models, i.e. the critical density $s_{\rm crit}$, the local star formation efficiency $\epsilon_\mathrm{loc}$, and the method to estimate turbulence. The following rows are the differences in the feedback subgrid models, namely the time sampling of the supernovae explosions and its boundaries, the average supernova rate (the number of supernovae per $100\rm\ M_\odot$), the method for the thermal energy injection, the equation for the momentum injected in the adiabatic phase of the supernova, the transition and the radiative phase, where this momentum is injected, and the metal yield of the ejecta.}
                \makebox[\textwidth][c]{
                \begin{tabular}{|p{3.8cm}cp{4.5cm}p{4cm}p{2.3cm}|}
                    \hline
                     & \vline & \textbf{\KI} & \textbf{\KR} & \textbf{\DC} \\
                    \hline
                    $s_{\rm crit}$ & \vline 
                        & $\ln \left[0.62 \alpha_{\rm vir} \mathcal{M}^{2}\right]$ 
                        & $\ln \left[\alpha_\mathrm{vir}\left(1+\frac{2\mathcal{M}^4}{1+\mathcal{M}^2}\right)\right]$ 
                        & $\ln \left[0.62 \alpha_{\rm vir} \mathcal{M}^{2}\right]$ \\ [.1 cm]
                    $\epsilon_\mathrm{loc}$ & \vline & 0.1425 & 0.5 & 0.1425 \\ [.1 cm]
                    Turbulence $\sigma_{1\mathrm{D}}$ & \vline & Local estimation & Subgrid turbulence model & Local estimation \\ [.1 cm]
                    Supernova sampling & \vline & Single explosion & Uniform time sampling & Single explosion  \\ [.1 cm]
                    Time of supernova & \vline & $10\rm\ Myr$ & $0$ to $20\rm\ Myr$ & $10\rm\ Myr$ \\ [.1 cm]
                    Average supernova rate & \vline & 2 SN$/100\rm\ M_\odot$ & 8 SN$/100\rm\ M_\odot$ & 2 SN$/100\rm\ M_\odot$ \\ [.1 cm] 
                    Thermal energy injection  & \vline & None & Random cell within the 8 parent cell & In the central cell + cooling turned off \\ [.1 cm]
                    Momentum in the adiabatic phase & \vline & $B_\mathrm{ph}\sqrt{2 \chi m_\mathrm{ej} f_\mathrm{e} E_\mathrm{SN}}$ & None & None \\ [.1 cm]
                    Transition & \vline 
                        & $\chi \equiv {\mathrm{d}m_{\mathrm{W}}}/{\mathrm{d}m_{\mathrm{ej}}}$ 
                        & $R_\mathrm{cool} = 43.6 n_\mathrm{H}^{-0.42} Z'^{-0.05}\rm\ pc$ & None \\ [.1 cm]
                    Momentum in the radiative phase $\rm [M_{\odot}\,km\,s^{-1}]$
                        & \vline & $2.5 \times 10^{5} N_\mathrm{SN}^{{16}/{17}} n_\mathrm{H}^{-{2}{/17}} Z'^{-0.14} \mathrm{e}^{-{\Delta x}/{R_\mathrm{st}}} + 5 \times 10^{5} N_\mathrm{SN}^{{16}/{17}} Z'^{-0.14} (1-\mathrm{e}^{-{\Delta x}/{R_\mathrm{st}}})$
                        & $2.66 \times 10^{5} \eta_R N_{\mathrm{SN}} Z'^{-0.114} n_\mathrm{H}^{-0.19}$ & None \\ [.1 cm]
                    Momentum injection sites & \vline & Neighbouring cells & Neighbour cells + source term & None \\ [.1 cm]
                    Metal yield & \vline & 7.5\% & 10\% & 7.5\% \\
                    \hline
                \end{tabular}
                }
                \label{tab:models_diff}
            \end{table*}

    \subsection{\DC\ simulation}  \label{subsec:DCsim}
        \subsubsection*{Star formation}
            We use the same star formation recipe as for \KI\ here.
        \subsubsection*{Feedback}
            We refer to this last model as \DC, as it follows the delayed cooling feedback model from \citet{Teyssier2013}. In this model, the supernova energy is injected in the form of thermal energy, and the overcooling problem is solved by inhibiting cooling locally. This prescription is used for example in the MIRAGE simulations \citep{Perret2014} as well as in many comparative studies of feedback models \citep{Dale2015, Gabor2016, Rosdahl2017, Nunez-Castineyra2021}, and similar implementations are commonly used in major simulation suites with other codes, such as NIHAO \citep{Wang2015}.

            As in the \KI\ simulation, star particles undergo all their supernovae explosions at once, $10\rm\ Myr$ after they were created. The total return mass and supernovae energy are deposited in the host cell with the same yield as in \KI. In order to prevent excessive cooling, cooling is turned off for a given amount of time after the supernova energy is injected.
            The duration depends on an additional variable $\epsilon_\mathrm{DC}$ which increases every time a supernova occurs. This quantity is advected with the gas and is only used as a passive scalar which is exponentially damped at each timestep with a characteristic dissipation timescale of $10\rm\ Myr$ \citep{Teyssier2013}.

            \vspace{0.3 cm}
            We summarise the main differences between \KI, \KR, and \DC\ in Table~\ref{tab:models_diff}. The primary distinction for star formation lies in the varying methods used to estimate the velocity dispersion $\sigma_{1\mathrm{D}}$. For feedback, major changes are the energies and supernova rates used in each model, the time sampling of the supernova (single explosion in \DC\ and \KI\, sampled in time in \KR), and how the supernova is modelled: mechanical feedback (\KI), thermal feedback combined with mechanical feedback (\KR) or delayed cooling (\DC). \KI\ also includes compensation for unresolved H$\,\textsc{ii}$ regions. The observed differences observed in the resulting galaxies might thus come from different implementations, but could also be due to variations in the parameter choices. As this paper aims to highlight differences between the models as a whole, we do not try to find the exact mechanism causing these differences.

\section{Global properties} \label{sec:glob_prop}

    Our objective is to create galaxies with similar stellar masses with the three different models and test whether these models produce different CGM properties. In this section, we first focus on the stellar properties of the calibrated galaxies, before comparing the global gas properties at halo scale.

    \subsection{Star formation histories}
        \begin{figure}
            \begin{subfigure}[h]{\columnwidth}
        	       \includegraphics[width=\columnwidth]{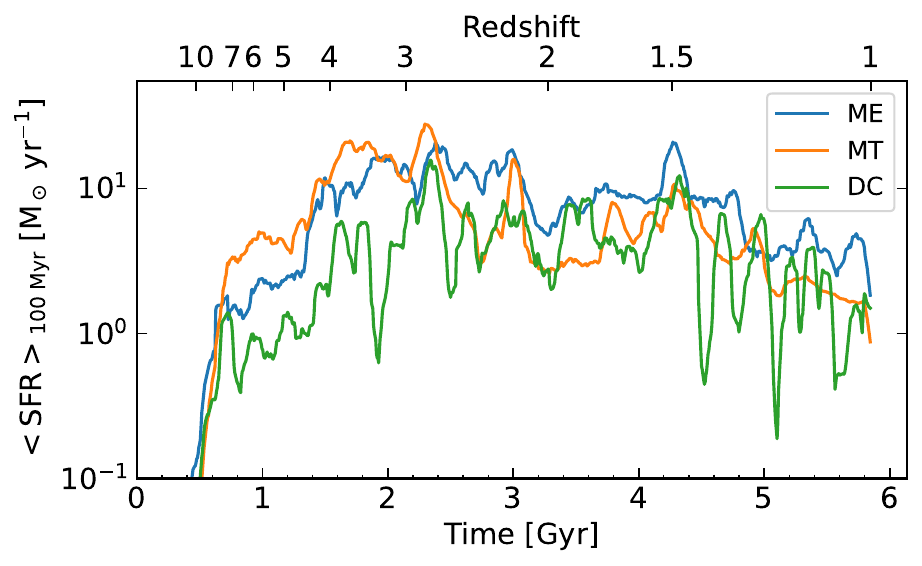}
            \end{subfigure}
            \begin{subfigure}[h]{\columnwidth}
        	       \includegraphics[width=\columnwidth]{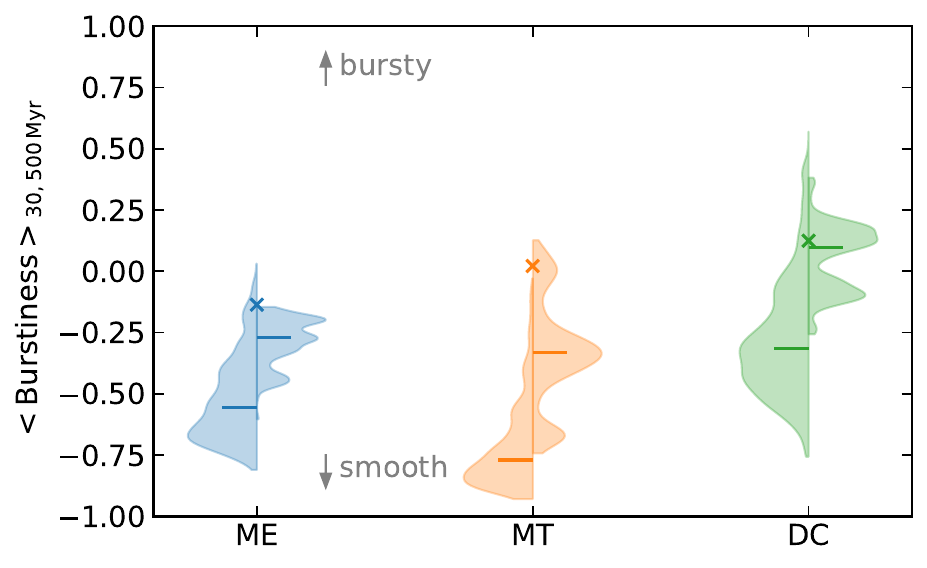}
            \end{subfigure}
            \caption{Star formation histories (top) for mechanical feedback (\KI, blue), thermal feedback combined with mechanical feedback (\KR, orange) and delayed cooling (\DC, green) and burstiness (bottom). The star formation history is computed using a $\rm100\ Myr$ moving average for clarity, and the burstiness is quantified through a $\rm1\ Myr$ moving average to highlight short-term fluctuations. Following Eq.~\ref{eq:burstiness}, we compute the burstiness from $\rm2\ Gyr$ to the end of the simulation (markers) over moving periods of $\rm30\ Myr$ and $\rm500\ Myr$ (left and right side of the violins). The violins show the probability density function (PDF) of the burstiness, and the horizontal bar corresponds to the median.}
            \label{fig:SFR}
        \end{figure}
        In the top panel of Fig.~\ref{fig:SFR}, we show the star formation rate as a function of time. We compute it from the age and initial mass of all stellar particles within the galaxy at the last snapshot, and smooth it over $100\rm\ Myr$. Here and for the rest of the paper, we define all stellar properties within $0.1\ R_{200}$.
        The three simulations have a comparable level of star formation at late times ($z<3$) but their star formation histories show distinct features. The three simulations start forming stars simultaneously, with an initial star formation rate sharply increasing. The three models exhibit contrasting behaviour, with \KR\ reaching $\approx3.4\rm\ M_\odot\,yr^{-1}$, and both \KI\ and \DC\ reaching $\approx1.6\rm\ M_\odot\,yr^{-1}$.
        The three simulations also reveal different levels of burstiness, with \DC\ showing the widest variations, in the form of deep dips where the star-forming activity drops by 1 order of magnitude on a short-but-significant timescale.
        Analysing the same models (albeit using feedback sampled in time with \KI) with much greater time resolution around a star formation burst in a similar simulation, \citet{Rey2022} found that stellar particles form through distinct pathways with different models. While stars form in highly supersonic environments due to turbulent compression in \KR, they form in mildly turbulent environments with much lower virial parameters in \KI\ and \DC\ (see Fig.~6.3 in \citealt{Rey2022}). 

        To quantify how the burstiness varies among the three simulations, we compute a burstiness parameter inspired by \citet{Goh2006}, as 
        \begin{equation} \label{eq:burstiness}
            B_{\Delta t}=\frac{\sigma_{\Delta t}/\mu_{\Delta t}-1}{\sigma_{\Delta t}/\mu_{\Delta t}+1},
        \end{equation}
        with $\mu_{\Delta t}$ and $\sigma_{\Delta t}$ the mean and standard deviation of the star formation rate over a time $\Delta t$. When the standard deviation is much smaller than the mean star formation rate ($\sigma_{\Delta t} \ll \mu_{\Delta t}$), the burstiness $B$ approaches $-1$, indicating no burstiness. Conversely, when the standard deviation is large, $B$ tends to $1$, representing maximal burstiness. We compute the mean and standard deviation of the star formation rate as 
        \begin{align}
            \mu_{\Delta t} &= \frac{1}{{\Delta t}} \sum\limits_{i\in[t\pm\Delta t/2]} \mathrm{SFR}_i^{1}, \\
            \sigma_{\Delta t}^2 &= \frac{1}{N} \sum\limits_{i\in[t\pm\Delta t/2]} (\mathrm{SFR}_i^{1}-\mu_{\Delta t})^2,
        \end{align}
        where $N$ is the number of bins over $[t\pm\Delta t/2]$, and $\mathrm{SFR}_i^{1}$ is the star formation rate smoothed with a $1\rm\ Myr$ sliding average. In all our computations, we restrict the star formation rates to $t>2\rm\ Gyr$, when the galaxy in \KI\ becomes more massive than 20\% of its final stellar mass.

        In the bottom panel of Fig.~\ref{fig:SFR}, we show the resulting estimate of burstiness for $\Delta t= 30, 500\rm\ Myr$ and over the whole simulation, with $t>2\rm\ Gyr$. Over the entire simulation (shown with crosses), we find a burstiness of $B\approx-0.14$ and $B\approx0.02$ for \KI\ and \KR, and $B\approx0.12$ for \DC, confirming the overall higher burstiness observed in the star formation rate with \DC. 

        To get a sense of how the burstiness differs, we additionally compare the burstiness over $\Delta t=500\rm\ Myr$, roughly encapsulating the large bursts observed in \DC, and a shorter timescale of $\Delta t=30\rm\ Myr$. The left (right) side of each violin represents the PDF of the burstiness over moving windows of $\rm30\ Myr$ ($\rm500\ Myr$). Focusing first on the larger bursts, we find that \KI\ and \KR\ exhibit a similar main peak of burstiness at $B_{500}\approx-0.3$. \KR\ exhibits another region of even lower burstiness, corresponding to times when the SFR is almost flat, such as over $3.2-3.8\rm\ Gyr$. \DC\ stands far above \KI\ and \KR\ with a positive median burstiness.
        Focusing on the smaller timescale ($\Delta t=30\rm\ Myr$), we find that the distribution is quite different and much lower for all simulations. Although the median burstiness in \KI\ decreases, \KI's burstiness distribution is less impacted by the change of timescale than the distributions of other models. \KR\ drops to much lower values. This possibly hints at supernovae disrupting star-forming clouds and hindering star formation faster in \KI\ than in \KR. This could be because energy injection is uniformly sampled up to $20\rm\ Myr$ in \KR, whereas in \KI, energy is injected instantaneously at $10\rm\ Myr$. It is also likely that the sampled injection times in \KR\ slow down star formation in the cloud so that it can form stars for a longer period before reaching a critical state and exploding. For \KI, there is no supernova feedback for the first $10\rm\ Myr$, and star formation can thus proceed very rapidly, producing after $10\rm\ Myr$ a large and synchronous number of supernovae events that disrupt the cloud. While the median burstiness in \DC\ also decreases, it remains far above other models. 
        We thus find very distinct behaviour in the three simulations. \DC\ is significantly more bursty than both other models at all timescales considered, a direct consequence of more disruptive feedback leading to frequent suppression of star formation. We also find contrasting behaviour between \KI\ and \KR. \KI\ exhibits a slightly lower burstiness on $30\rm\ Myr$ timescales compared to $500\rm\ Myr$ timescales, whereas in \KR\ the difference is much more pronounced. This highlights the distinct star formation regulation in the three models, and the role of both star formation and feedback subgrid models, although we see with \DC\ that feedback plays a critical role.

        \begin{figure}
           \includegraphics[width=\columnwidth]{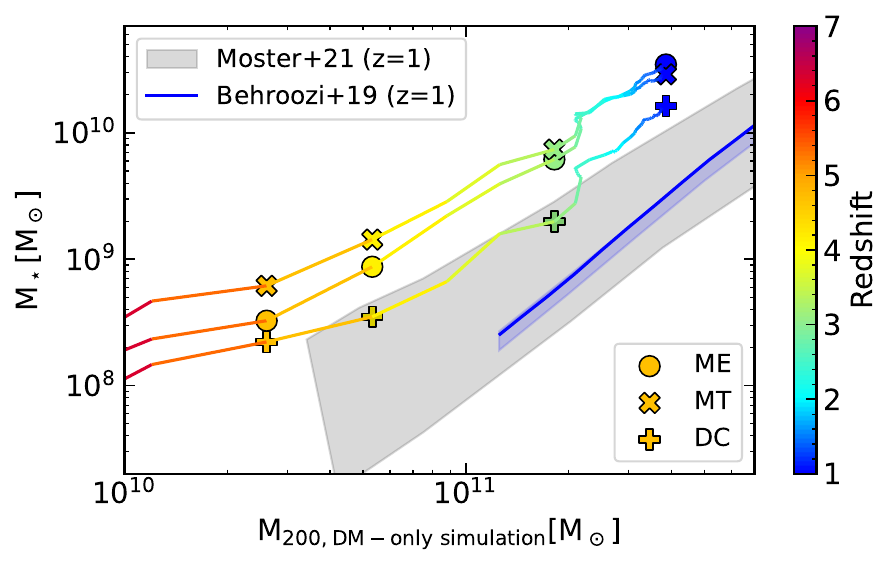}
            \caption{Stellar mass-to-halo mass relation. Each marker represents a different simulation, as indicated in the legend. The colour code shows the redshift, which goes down to $z=1$ (blue). The solid blue line corresponds to the empirical model from \citet{Behroozi2019}, with the blue shaded area showing the $16^\mathrm{th}-84^\mathrm{th}$ percentiles range. The grey shaded area indicates the broad contour of predictions of \citet{Moster2021} through reinforced learning\protect\footnotemark. Both predictions are at redshift $z=1$.}
            \label{fig:Mstar_Mhalo}
        \end{figure}
        \footnotetext{The strongest constraint to connect properties of galaxies and dark matter haloes is the peak halo mass and not the current halo mass. Thus, the y-axis in \citet{Behroozi2019} and \citet{Moster2021} is the peak halo mass. In our simulations, there are no events such as major mergers that can remove a significant fraction of the halo mass. We thus assume both masses to be identical.}
        In Fig.~\ref{fig:Mstar_Mhalo}, we show the stellar mass-to-halo mass relation for our three simulations. The halo mass is measured from a twin dark-matter-only simulation to match the approach of \citet{Behroozi2019} and \citet{Moster2021} since it can vary with baryonic physics \citep{Velliscig2014}.
        Due to stronger feedback, the stellar mass remains roughly twice lower in \DC\ than in the other two models down to the end of the simulations at $z=1$. Near redshift $z=2.5$, numerous mergers in rapid succession lead to a quick increase in the stellar mass of the galaxy in all simulations and results in the step that can be seen at a halo mass of $M_\mathrm{halo, 200}\approx2-3\times 10^{11}\rm\ M_\odot$.

        Our simulations are above the stellar mass-to-halo mass mass relation inferred from observations at $z=1$. As galaxies get more massive, their gravitational potential becomes too strong for the supernovae feedback models to regulate star formation, and AGNs (which are not simulated here) may take over as the driver of feedback to solve this problem \citep{Silk1998, Somerville2015, Naab2017}. Alternatively, different star formation models leading to bursty star formation may produce stronger outflows that inhibit galaxy growth \citep{Kang2025}. The high stellar mass might be explained by the history of the simulated galaxy, as we only follow a single set of initial conditions, or by the stochasticity of galaxy formation \citep{Keller2019, Genel2019}. Lastly, some uncertainties still weigh on these relations, as we can see with the large error bars from \citet{Moster2021}. Sources of such uncertainties include underestimating the observed stellar mass \citep{Munshi2013}. 
        \vskip 0.2cm

        We saw that our three simulations produce galaxies with similar stellar masses, slightly above the stellar mass-to-halo mass relation inferred from observations. The \KI\ and \KR\ runs yield very similar masses, despite slightly different, moderately bursty star formation histories. The \DC\ model has very strong feedback which shuts down star formation over a significant fraction of the time, resulting in a galaxy about half the mass of the other runs. These properties being understood, we now move on to analyse the properties of the gas in these three simulations.

    \subsection{The multiphase nature of ISM and CGM gas}
        \begin{figure}
            \centering
            \includegraphics[width=\columnwidth]{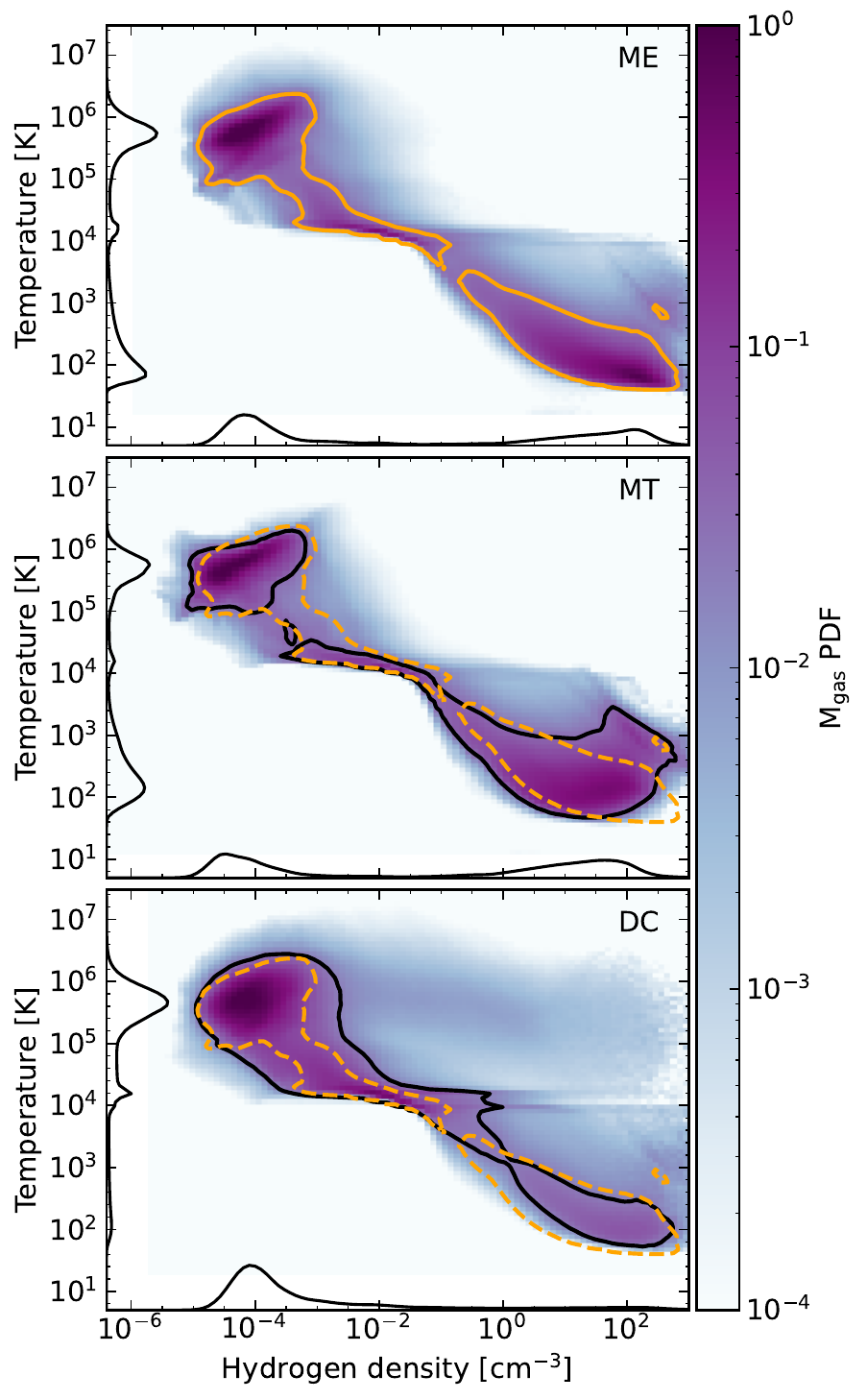}
            \caption{
            Phase diagrams of the gas contained within $R_{200}$ for each simulation. The result consists in 100 snapshots stacked over $1\rm\ Gyr$ from $z=1.3$ to $z=1$. From top to bottom are \KI, \KR, and \DC\ with a colour code corresponding to the hydrogen mass normalised probability density function. The solid line contour encompasses 90\% of the mass within $R_{200}$, and the outline in the orange dotted line is that of \KI.
            We show the mass-weighted temperature and density probability distribution functions on the left and lower sides of each panel.}
            \label{fig:PD_models}
        \end{figure}

        In Fig.~\ref{fig:PD_models}, we compare the density-temperature distributions of the gas in the dark matter haloes of \KI\ (top), \KR\ (middle), and \DC\ (bottom). Each panel shows the stacked phase diagram of 100 outputs from redshift $z = 1.3$ down to redshift $z=1$.
        The gas in all simulations is split into three main phases. The first phase is the hot ($10^5 - 10^6\rm\ K$) and low-density ($10^{-5} - 10^{-3}\rm\ cm^{-3}$) shock-heated gas. At higher densities, cooling becomes efficient, and the gas quickly drops to $10^4\rm\ K$ to develop the second phase of the density-temperature diagram in equilibrium between cooling and photoionization heating from the UV background. This gas is spread over a wide range of densities. At higher densities, cooling becomes rapid again and is responsible for the third phase: dense ($30 - 300\rm\ cm^{-3}$) and cold gas ($10^2 - 10^3\rm\ K$) in which star formation can occur. For the rest of the paper, we categorize the gas into three phases based on temperature: cold ($T < 10^{3.5}\rm\ K$), warm ($10^{3.5} \rm\ K < T < 10^{4.5}\rm\ K$), and hot ($T > 10^{4.5}\rm\ K$), corresponding to the three distinct peaks observed in the phase diagrams.

        We find a striking difference between \DC\ and the two other models, as gas in \DC\ populates the intermediate to high-density ($n_\mathrm{H} \gtrsim 10^{-2}\rm\ cm^{-3}$), high-temperature ($T \gtrsim 10^5\rm\ K$) part of the diagram where cooling is supposed to be highly efficient. This is a direct consequence of the delayed cooling model, and the thermal state of these cells should be taken with caution. We have checked that the mass (volume) of gas in these cells represents less than 0.1 per cent of the total gas mass (volume) within $0.1\ R_{200}$ at $z=1$ and much less in the CGM. This hot dense gas is concentrated around stars and does not affect our results in the next sections. However, it might play a role in star formation itself by keeping gas from cooling and forming a new generation of stars.

        Focusing now on the density distribution (the solid black lines at the bottom of each panel), we find that \DC\ exhibits a notable difference relative to the other simulations. For both \KI\ and \KR, the density distribution has two peaks, roughly containing an equal mass of gas. These peaks respectively correspond to gas in the CGM (largely hot and very diffuse) and gas in the ISM (predominantly cold and relatively dense). This starkly contrasts with the results of \DC, which has most of its gas in the low-density, high-temperature CGM. In this simulation, the ISM contains much less gas than in the other two models, which is another manifestation of the very efficient (and ejective) feedback obtained with delayed cooling.

        In the ISM, a large fraction of gas is expected to be in the warm phase \citep{Ferriere2001, Troland2004}. Recent high-resolution simulations focused on idealised galaxies \citep[e.g.,][]{Katz2022} or the solar neighbourhood \citep[e.g.,][]{Kim2017} have observed comparable distribution. In contrast, the mass fraction of the warm phase within the ISM over $z = 1.3-1$ in \KI\ and \KR\ is $\approx5\%$. The cold phase dominates the ISM gas with a mass fraction of $\approx94\%$. Conversely, the cold and warm gas phases over that period respectively represent 49\% and 31\% of the ISM mass in \DC. Although the mass distribution across the different phases remains stable over that period in \KI\ and \KR, it largely fluctuates in \DC\ with the cold and warm gas mass fractions oscillating between 18-66\% and 20-54\%. The hot gas mass fraction in \DC\ can reach up to 60\%, whereas it stays below 2\% in \KI\ and \KR\ throughout the entire period. This behaviour is anticipated and is a direct result of delayed cooling, which rapidly heats large regions of gas that do not cool down quickly.
        Despite the similar stellar masses of the galaxies in our three simulations, the varying subgrid models for feedback and star formation lead to distinct differences in the gas content of the galaxy and its surrounding halo.

    \subsection{Outflow properties}
        \begin{figure}
            \begin{subfigure}[h]{\columnwidth}
        	       \includegraphics[width=\columnwidth]{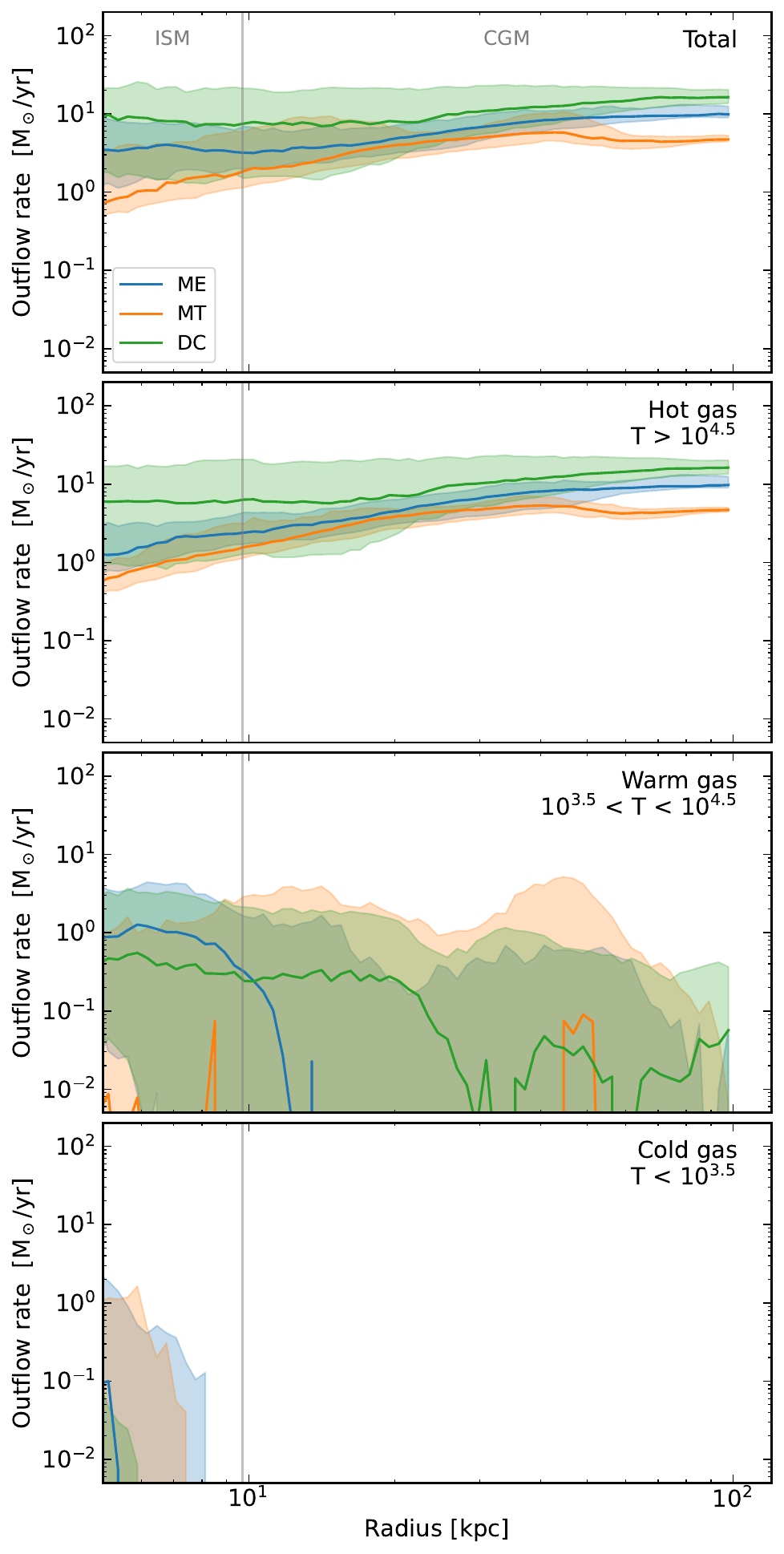}
            \end{subfigure}
            \caption{Spherical instantaneous outflow rates as a function of radius in the CGM of \KI, \KR, and \DC. We show total outflow rates (top) and split it between hot (T$>10^{4.5}\rm\ K$), warm ($\rm 10^{3.5}<T<10^{4.5}\rm\ K$), and cold (T$<10^{3.5}\rm\ K$) outflow phases (from top to bottom). The solid lines correspond to the median of each shell over $1\rm\ Gyr$ from $z=1.3$ to $z=1$, and the shaded area corresponds to the 15.9 and 84.1 percentiles. We split the figure between the ISM and the CGM by placing a vertical grey line at $0.1\ R_{200}$.}
            \label{fig:radial_outflows}
        \end{figure}

        Having understood how the properties of the ISM and CGM differ in each simulation (i.e. stellar properties, temperature), we investigate the main culprit in driving these properties: galactic outflows. At a given radius $r$, we compute the instantaneous outflow rates using a method similar to \citet{DallaVecchia2008} and \citet{Nelson2019b}, as $$ \dot{M}(r) = 4\pi r^2 \frac{ \sum_{i}^{N_\mathrm{shell}} (m_i v_{r, i})} {\sum_{i}^{N_\mathrm{shell}} \Delta x_i^3},$$ with $4\pi r^2$ the surface of the spherical shell at a distance $r$ from the galaxy centre, and $v_{r, i} = {\mathbf{v}_i \cdot \mathbf{r}_i}/{|\mathbf{r}_i|}$ the radial velocity of the gas, where the velocity is obtained by subtracting the mean motion of the dark matter halo. The two summations are done over $N_\mathrm{shell}$ cells within a shell defined by $r_s\in[r\pm\Delta x_i/2]$. $\Delta x_i$,  $m_i$, $\mathbf{r}_i$, and $\mathbf{v}_i$ respectively describe the cells' size, mass, position, and velocity in Cartesian coordinates. To minimize errors due to Cartesian grid structures, we compute the outflows using the factor $\dot{M}(r) = {4\pi r^2 }/{ \sum_{i}^{N_\mathrm{shell}} \Delta x^3}$ instead of simply dividing the total momentum by the thickness of the shell with $1/\Delta x$. Additionally, we exclude all cells within $0.2\ R_{\rm vir, sub}$ of any subhaloes containing stars, with $R_{\rm vir, sub}$ defined as the virial radius for subhaloes computed by the ADAPTAHOP halo finder \citep{Tweed2009}. Lastly, to avoid counting turbulent motions, we restrict data to cells with radial velocities ${v_{r, i}>0.5v_\mathrm{circ}}$ for outflows (${v_{r,i}<-0.5v_\mathrm{circ}}$ for inflows), with $v_\mathrm{circ} = \sqrt{\frac{G\rm M_{halo}}{R_{200}}}$, and $\rm M_{halo}$ the total mass within the halo. Over $z=1-1.3$, ${128\rm\ km/s<v_\mathrm{circ}<139\rm\ km/s}$.

        In Fig.~\ref{fig:radial_outflows}, we show the outflow rates of the three simulations as a function of radius. 
        The three models produce total outflow rates within 1 dex of each other, with \DC\ higher than \KI, which is higher than \KR. For all three models, the CGM mass flux increases with radius, a sign of outflows entraining CGM gas, as explained in \citet{Mitchell2020}. For \KR, \KI, and \DC, the outflow rates respectively range from 2, 3.5, and $8\rm\ M_\odot/yr$ at $0.1\ R_{200}$ to 7, 14, and $24\rm\ M_\odot/yr$ at $\ R_{200}$, revealing a factor of three to four increase in flux, the highest increase being seen in \KI\, and the smallest in \DC. By the time the outflows reach the halo edge, at least 2/3 of the gas is entrained from the CGM. 

        We now focus on the different phases of the outflows and find that in all three simulations, the total outflow rates are widely dominated by hot gas. The entrained gas is hence mainly hot gas from the CGM.
        In warm gas, we find contrasting behaviour between the three models. In the CGM, \DC\ exhibits continuous and overall higher outflow rates than \KI\ and \KR, consistent with time. This warm outflowing gas is relatively conserved up to $\approx25\rm\ kpc$, from whence its median value start decreasing. Conversely, \KI\ and \KR\ vary significantly over time (see the shaded area), with the median in \KI\ dropping sharply in the CGM and \KR\ showing a single peak of outflowing gas at $\approx 0.5 R_{200}$. This highlights the different behaviour of warm outflows in the three simulations. In \DC, warm outflows are continuously driven in time, while in \KI\ and \KR, warm outflows are episodic, oscillating between states in which the galaxy is sending plumes of hot gas propagating outwards and having low warm outflow rates.
        Lastly, there is no outflowing cold gas in any of our simulations. 

        Several recent studies suggest that hot outflows entrain a population of cold clouds from the ISM \citep{Tan2024} which then grow \citep{Gronke2023} at a rate set by mixing \citep{Tan2021} through cooling-induced pulsation \citep{Gronke2020b} if they manage to survive. Our result possibly shows how cold gas fails to be entrained as it is heated up while being ejected out of the galaxy and thermally mixed with hotter gas. The current main explanation in the literature for this inability to entrain cold gas into the CGM is a lack of resolution. The resolution in our CGM is at kiloparsec scale, while \citet{Gronke2018, Gronke2022} suggest a need for resolution better than $100\rm\ pc$, which corresponds roughly to the estimated size of CGM cold clouds \citep{McCourt2018}. Alternatively, non-thermal support provided by cosmic rays \citep{Salem2016, Ji2020} or magnetic fields \citep{McCourt2015, Ramesh2023b, Ramesh2024c} could explain the lack of cold gas outflows and survival. However, cosmic ray properties are still poorly constrained \citep{Ruszkowski2017, Hopkins2020b, Hopkins2021}, and their role is greatly affected by the parameters used \citep{Butsky2022, DeFelippis2024}. The impact of magnetic fields on the CGM also remains debated \citep{Li2020, Das2024b}.

        \begin{figure}
            \begin{subfigure}[h]{\columnwidth}
        	       \includegraphics[width=\columnwidth]{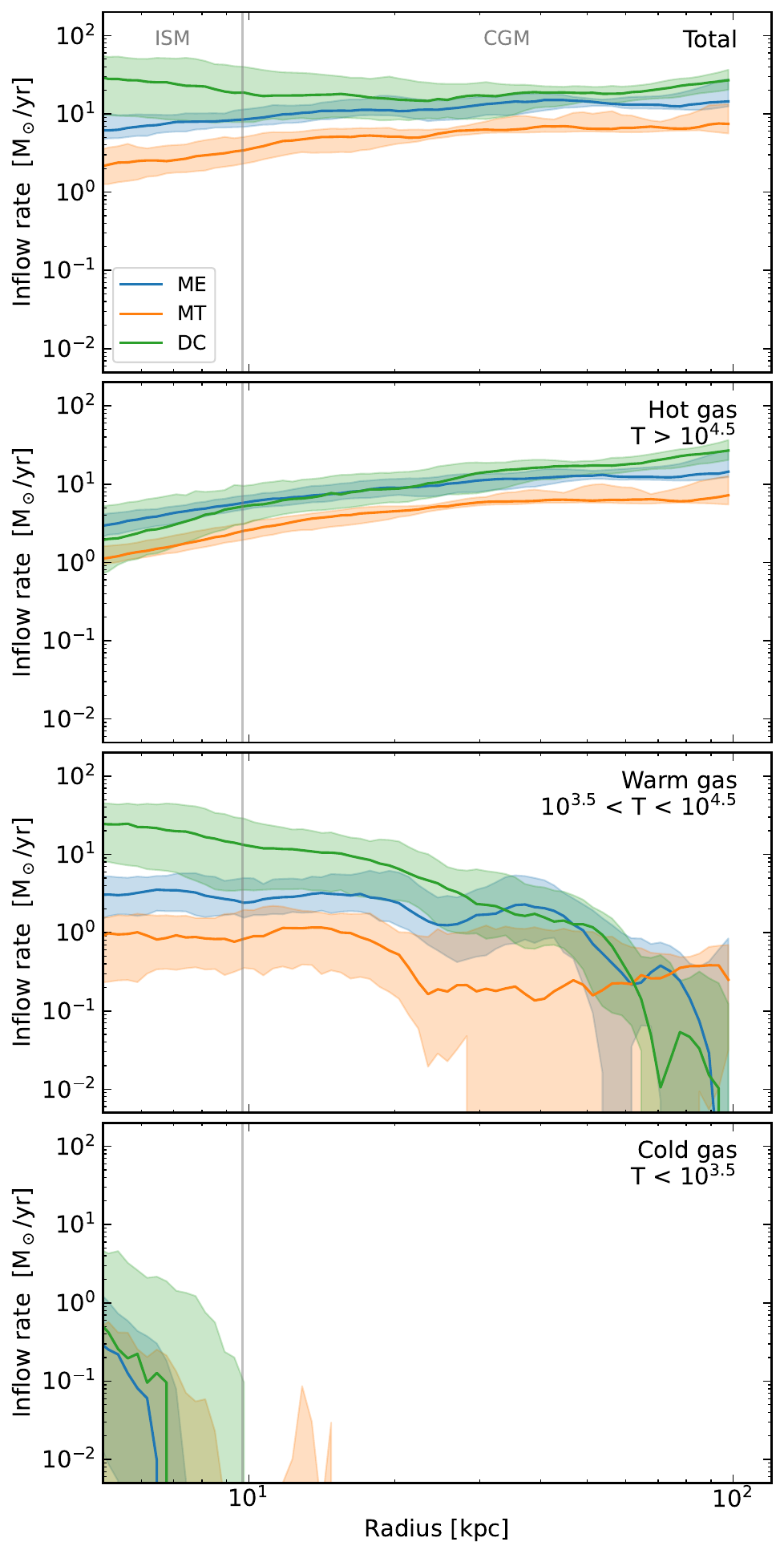}
            \end{subfigure}
            \caption{Spherical instantaneous inflow rates as a function of radius in the CGM of \KI, \KR, and \DC. We show total inflow rates (top) and split it between hot (T$>10^{4.5}\rm\ K$), warm ($\rm 10^{3.5}<T<10^{4.5}\rm\ K$), and cold (T$<10^{3.5}\rm\ K$) inflow phases (from top to bottom). The solid lines correspond to the median of each shell over $1\rm\ Gyr$ from $z=1.3$ to $z=1$, and the shaded area corresponds to the 15.9 and 84.1 percentiles. We split the figure between the ISM and the CGM by placing a vertical grey line at $0.1\ R_{200}$.}
            \label{fig:radial_inflows}
        \end{figure}

        \begin{figure*}
            \centering
            \begin{subfigure}[b]{\textwidth}
                \includegraphics[width=\textwidth]{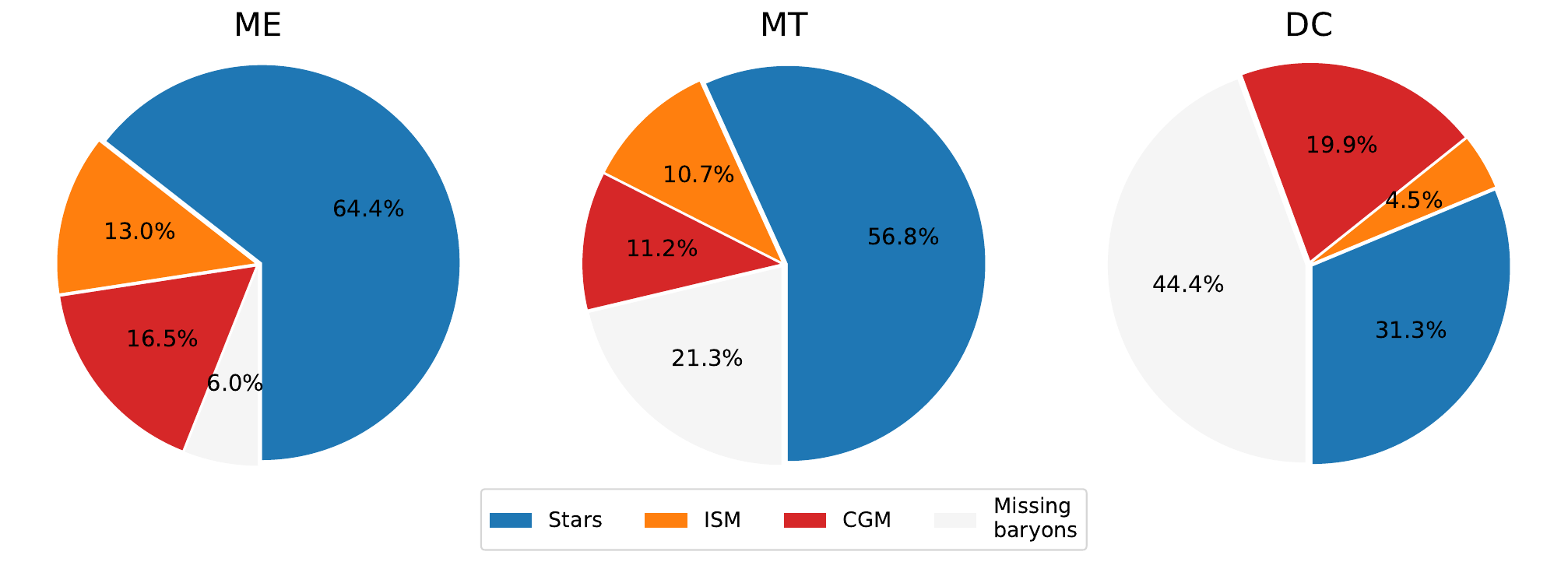}
            \end{subfigure}
            \caption{Fraction of baryons in different media of the simulation for \KI\ (left), \KR\ (centre), and \DC\ (right).
            The total baryonic mass considered is given by $f_\mathrm{b}\rm M_{200, min}$, with $\rm f_b\approx 0.154$ and $\rm M_{200, min}$ the dark matter mass within $R_\mathrm{200, min}$. 
            We split it into the stellar component (blue), the ISM (orange), the CGM (red) and the difference with the total (white). The difference between the baryonic mass of the simulation and that expected from the dark matter halo mass corresponds to the \textit{missing baryons}. The result consists of 100 snapshots stacked over $1\rm\ Gyr$ from $z=1.3$ to $z=1$. For each simulation, we use the smallest $R_\mathrm{200}$ among the three simulations $R_\mathrm{200, min}$ to define the different regions at each timestep. The ISM contains all the gas within $0.1\ R_\mathrm{200, min}$, and the CGM corresponds to gas between $0.1\ R_\mathrm{200, min}$ and $R_\mathrm{200, min}$.}
            \label{fig:PIES_baryons}
        \end{figure*}

        In Fig.~\ref{fig:radial_inflows}, we show the inflow rates of the three simulations. At first glance, we find once again total inflow rates within 1 dex of each other among the three simulations, with \DC\ once again exhibiting higher fluxes than \KI, which in turn are higher than \KR. However, there are differences between the three models. The total inflow rate in \KR\ decreases by a factor slightly higher than two from $\ R_{200}$ to $0.1\ R_{200}$, going from $7.5\rm\ M_\odot/yr$ to $3.5\rm\ M_\odot/yr$. In contrast, \DC\ shows the smallest relative decrease in total inflow rates with decreasing radius, going from $27\rm\ M_\odot/yr$ to $18.5\rm\ M_\odot/yr$, although the absolute decrease is the highest. This highlights how infalling gas in \KR\ is more efficiently prevented from falling onto the galaxy, possibly stopped by outflows. By looking at the inflow temperature decomposition, we find that the hot inflowing phase fully drives this behaviour. This is expected as stellar feedback and gravitational shock heat up the halo gas \citep{ Birnboim2003, Dekel2006, Ocvirk2008}.
        From $R_{200}$ to $\approx 5\rm\ kpc$, the inflow rate in \DC\ appears relatively constant. In reality, the inflow rates decrease down to $\approx25\rm\ kpc$, from which they increase again to reach the same value as measured at $R_{200}$. It means that at large radii the gas is indeed being prevented from falling, but below $\approx25\rm\ kpc$, there is added gas falling on the galaxy, likely in the form of recycled gas. By looking at the metallicity of the inflowing gas (not shown), we confirm that gas is being recycled as its metallicity increases with decreasing radius. By looking at the temperature of the inflows, we find that in \DC, inflow rates are dominated by hot gas in the outer CGM and warm gas in the inner CGM. It thus seems that hot gas in \DC\ is also being prevented from falling into the galaxy while warm gas is being recycled into the galaxy. This is consistent with the warm outflows observed at $\approx25\rm\ kpc$, where warm gas is either slowed down or transitions to a hotter phase. 
        In \KI, the inflow rate increases from $R_{200}$ to $\approx 0.5R_{200}$ possibly showing gas recycled at large radii, which is then conserved as falls into the galaxy. In \KR, the warm inflowing rate decreases with decreasing radius down to $\approx25\rm\ kpc$, where a sudden five-fold increase occurs. The warm inflow rate is then conserved at lower radii and exhibits a much lower time variability, possibly hinting at more consistent recycling flows. However, as the warm inflow rates of \KI\ and \KR\ are only a fraction of their respective total inflow rate, we cannot conclude what causes this behaviour. It might indeed show recycling, but it could also correspond to a fraction of the hot inflows being converted into warm gas, or more coherent inflows.
        Lastly, the inflow rates of cold gas are negligible in all three simulations, showing that the cold gas that fuels star formation in the galaxy is either cold gas infalling at velocities below $0.5v_\mathrm{circ}$, cold gas accreted through satellite mergers, or warm recycled gas cooling down once in the galaxy. By looking at inflows without removing the satellite contribution and without a velocity cut, we confirm that cold gas falls onto the galaxy from satellites, but there is no slow-moving infalling cold gas. We also compare both inflow and outflow rates of all gas phases without any velocity cut. As the velocity cut reduces the flow rates similarly in all three simulations, our conclusions on their comparative behaviour hold. However, while the outflow rate in \DC\ slightly decreases (as all models do in the other gas phases), it extends to somewhat larger radii in \KI. In \KR, the outflow rates increase and become similar to \KI, rapidly decreasing and extending to $\approx25\rm\ kpc$. This is consistent with the picture drawn from Fig.~\ref{fig:radial_inflows} in which warm outflowing gas slows down as it extends and is eventually recycled into the galaxy.

        In summary, both outflows and inflows are highly dominated by hot gas, except for inflowing gas in \DC\ which is dominated by warm gas in the inner CGM. For all phases, \DC\ exhibits the highest flux in the CGM while it is lowest in \KR. In all three simulations, hot outflows entrain hot gas from the CGM, while inflows are being prevented from infalling onto the galaxy. Warm outflowing gas in \KI\ and \KR\ exhibit very sporadic outflows. In \DC, warm outflows are stable through time and decrease as a function of radius as they get recycled and fall back onto the galaxy.

    \subsection{The baryon distribution} \label{sec:PIES}


        In Fig.~\ref{fig:PIES_baryons}, we now investigate the location of the baryons \textit{expected} within the halo, given by $f_\mathrm{b}\rm M_{200, min}$, with $\rm f_b \approx 0.154$ the universal baryon fraction, and $\rm M_{200, min}$ the dark matter mass within $R_\mathrm{200, min}$. We define $R_\mathrm{200, min}$ as the smallest $R_\mathrm{200}$ among the three simulations. We stack outputs over the redshift range $z=1-1.3$, and split $f_\mathrm{b}\rm M_{200, min}$ between stars, the ISM, the CGM, and the missing baryons (i.e. the difference between the expected amount of baryons and the sum of stars, ISM and CGM masses).
        This figure gives a visual impression of marked differences between the three simulations. Feedback processes reduce the baryon content to $\approx 56\%$ of the universal fraction in the \DC\ run, $\approx 79\%$ in \KR\ and $94\%$ in the \KI\ simulation. \KI\ and \KR\ have comparable distribution of gas fractions in the halo components, with most of the baryons in stellar form, and a similar fraction of CGM and ISM gas. Conversely, \DC\ has a much reduced stellar and ISM fraction and a CGM gas fraction higher than the other two simulations.

        \begin{figure}
            \centering
            \begin{subfigure}[b]{\columnwidth}  
                \includegraphics[width=\columnwidth]{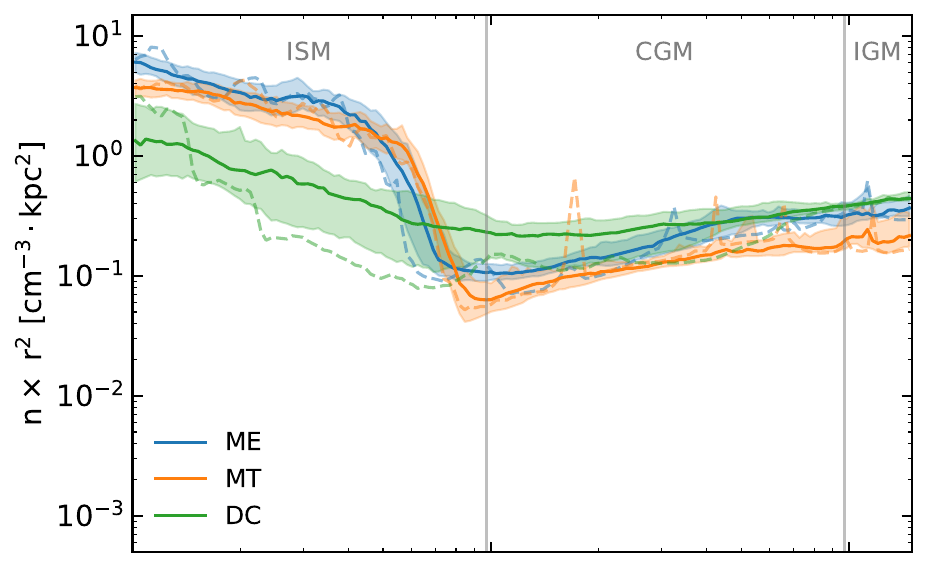}
            \end{subfigure}
            \begin{subfigure}[b]{\columnwidth}   
                \includegraphics[width=\columnwidth]{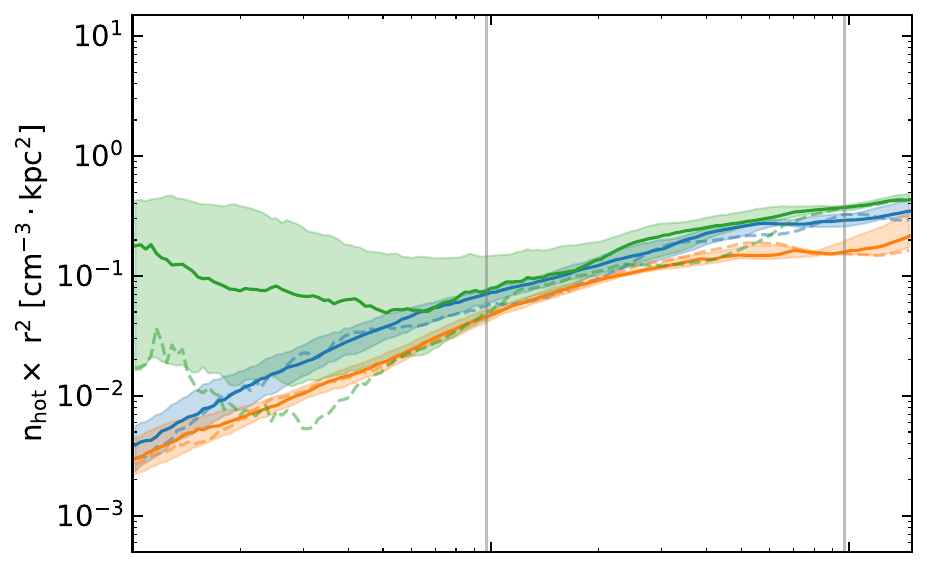}
            \end{subfigure}
            \begin{subfigure}[b]{\columnwidth}  
                \includegraphics[width=\columnwidth]{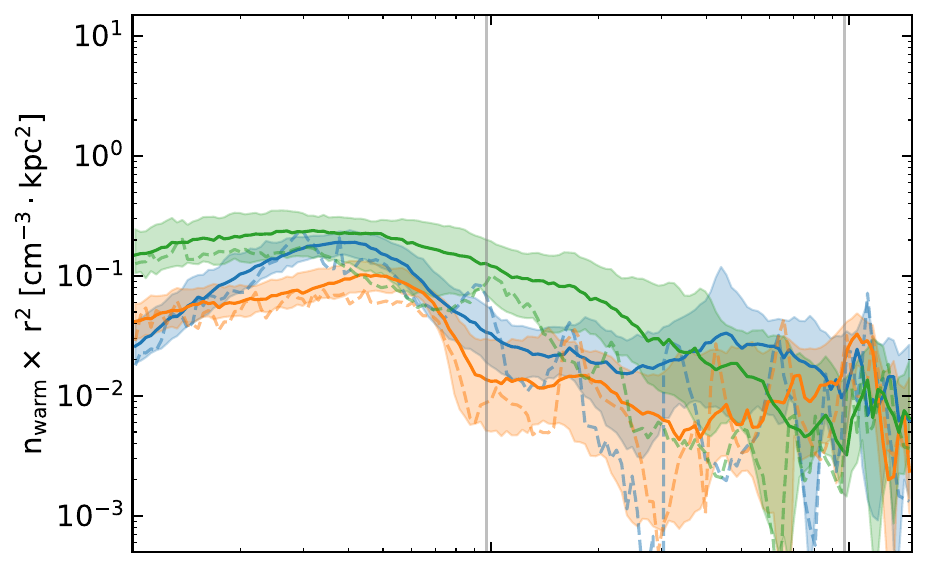}
            \end{subfigure}
            \begin{subfigure}[b]{\columnwidth}   
                \includegraphics[width=\columnwidth]{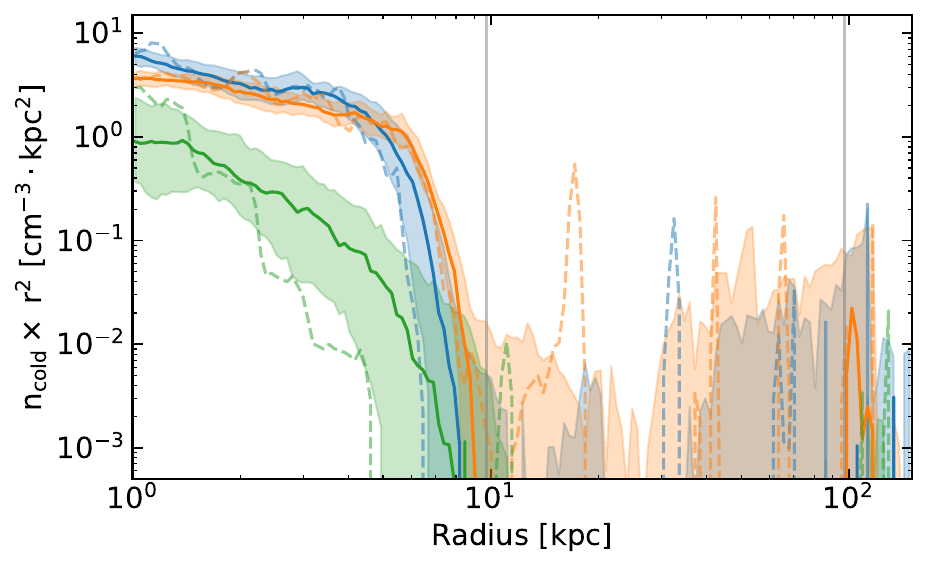}
            \end{subfigure}
            \caption{Radial density profile of all the gas in the halo (top), as well as the hot, warm and cold phases (from top to bottom) in \KI\ (blue), \KR\ (orange) and \DC\ (green). The solid lines correspond to the median of each shell over $1\rm\ Gyr$ from $z=1.3$ to $z=1$, and the shaded area corresponds to the 15.9 and 84.1 percentiles. We also show with dashed lines the radial density profile of the last snapshot, at $z=1$. Lastly, the figure is split between the ISM, the CGM and the IGM with two grey lines placed at $0.1\ R_{200}$ and $R_{200}$.}
            \label{fig:radial_nH}
            \vspace{-4pt} 
        \end{figure}

        In Fig.~\ref{fig:radial_nH}, we look at the radial density distribution of gas in the three simulations. We weigh the y-axis by the radius squared to compare the relative mass fractions. In \KI\ and \KR, we find densities of $\approx5\rm\ cm^{-3}$ in the ISM (at $1\rm\ kpc$), which then decrease sharply by 4 orders of magnitude, reaching densities of $\approx6-10\times10^{-4}\rm\ cm^{-4}$ at $0.1\ R_{200}$. As expected from Fig.~\ref{fig:PIES_baryons}, the ISM gas densities are lower in \DC, with no steep decrease at the galaxy edge, decreasing by slightly less than three dex over the same range, starting from $\approx1\rm\ cm^{-3}$. The transition between the ISM and the CGM gas profile is more continuous in \DC\ than in \KI\ and \KR\ and is caused by \DC's feedback removing gas efficiently from the ISM and ejecting it into the CGM and beyond, as observed in Fig.~\ref{fig:PIES_baryons}. In the CGM, the total gas density of the three models decreases similarly, roughly following an inverse square law with radius.

        Focusing on cold gas (lower panel of Fig.~\ref{fig:radial_nH}), we first find that \DC\ exhibits much lower densities than both \KI\ and \KR, which both have comparable density profiles. However, all three simulations have a negligible amount of cold gas in the CGM as the amount of cold gas drops at $0.1\ R_{200}$. The percentile distribution shown by the shaded area highlights short-lived cold gas in the form of satellites falling onto the main galaxy. We highlight their presence through the dashed lines, showing the density profile of a single snapshot. This is consistent with the absence of outflowing and inflowing cold gas observed in the CGM. As mentioned earlier, this possibly highlights how \textit{all} models fail to entrain cold gas out into the CGM.

        For warm gas, \DC\ exhibits the highest densities up to $\approx30\rm\ kpc$. In the ISM, we find that the warm gas content in \DC\ is quite significant and comparable to the cold gas content. In the CGM, the warm gas radial density also roughly evolves at a rate similar to the inverse square of the radius for \KI\ and \KR, but does so more steeply in \DC. We have seen in Fig.~\ref{fig:radial_outflows} and Fig.~\ref{fig:radial_inflows} that \DC\ shows signs of warm gas recycling in the CGM with outflows and inflows both decreasing as a function of radius. This leads to a larger amount of hot gas in the inner CGM and a steeper decrease in the warm gas radial density profile with increasing radius.

        Lastly, \DC\ shows higher ISM densities in the hot phase than the other models. We have seen in Fig.~\ref{fig:PD_models} that gas in \DC\ unrealistically populates the high-density ($n_\mathrm{H} \gtrsim 10^{-2}\rm\ cm^{-3}$), high-temperature ($T \gtrsim 10^5\rm\ K$) part of the halo phase diagram. This gas is close to star formation sites in which density is still high and where cooling has been turned off when supernova explosions are initiated. Thus, although this gas represents a small fraction of the total gas mass, it is expectedly dominant in radial density plots focusing on the hot gas phase. In the CGM, all three models show similar behaviour, with a decrease that scales inversely with radius. Outflows in all three models are dominated by hot gas, carrying it further than warm and cold gas. As a result, hot gas dominates the halo and displays the shallowest radial density decline among the three gas phases. However, note that while feedback can temporarily affect the pressure equilibrium of the halo, the radial density profile of the hot gas is constrained by the requirement that the total pressure gradient (bulk flows + turbulence + thermal) needs to balance gravity.

    \subsection{The metal budget} 
        \begin{figure}
            \begin{subfigure}[h]{\columnwidth}
        	       \includegraphics[width=\columnwidth]{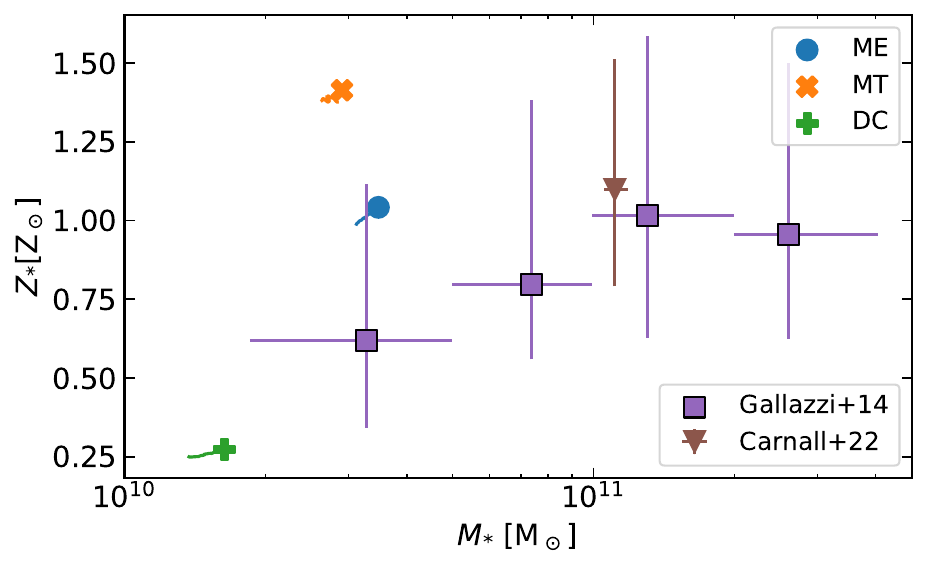}
            \end{subfigure}
            \caption{Mass-weighted stellar metallicity relation for \KI, \KR, and \DC, shown with markers. The lines associated with the markers show the metallicity evolution from $z=1.3$ to $z=1$. We compare our results to \citet{Gallazzi2014} at $0.65\leq z \leq0.75$ and \citet{Carnall2022} at $1<z<1.3$.}
            \label{fig:MZRs}
        \end{figure}

        \begin{figure*}
            \centering
            \begin{subfigure}[b]{\textwidth}
                \includegraphics[width=\textwidth]{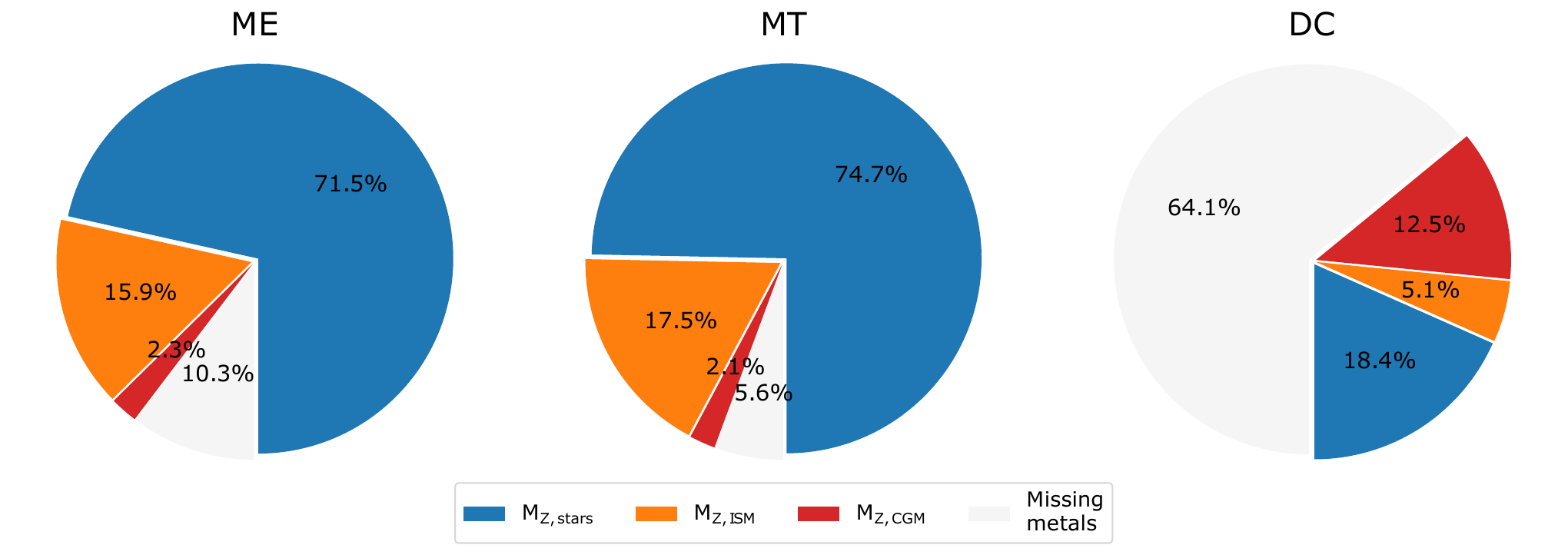}
            \end{subfigure}
            \caption{Fraction of metals in different media of the simulation for \KI\ (left), \KR\ (centre), and \DC\ (right).
            The total mass of metals considered here is the mass of metals created through the simulation by the stars within the halo at each timestep. We split it into the stellar component (blue), the ISM (orange), the CGM (red) and the difference with the total (white). This difference corresponds to the fraction of metals ejected out of the halo. The ISM, CGM, and $R_\mathrm{200, min}$ are defined as in Fig.~\ref{fig:PIES_baryons}. The result consists of 100 snapshots stacked over $1\rm\ Gyr$ from $z=1.3$ to $z=1$.}
            \label{fig:PIES_metal}
        \end{figure*}

        Having looked at the density properties of the halo, we now turn to metallicity. We show in Fig.~\ref{fig:MZRs} the stellar mass metallicity relation (MZR) over the redshift range $z=1-1.3$. The stellar metallicity is mass-weighted, and computed within $0.1\ R_{200}$. We find that \KI\ and \DC\ are both compatible with observations. As the galaxies have similar stellar masses, they should have produced a comparable amount of metals. However, this is not the case for the \KR\ model, which uses a higher metal yield than \KI\ and \DC\ (0.1, compared to 0.075). This leads to a stellar metallicity much higher in \KR\ than in \KI, above observational constraints. There is however a factor of four difference between the stellar metallicity in \KI\ and \DC, which is not explained by a different yield factor. This difference is due to the feedback models themselves, showing us that they regulate metals differently. \DC\ is very efficient in removing metals from the galaxy and the halo, while \KI\ and \KR\ both retain most of the metals locked in stars.


        \vspace{0.3 cm}
        For similar halo masses as ours, the \textsc{EAGLE} and the \textsc{SIMBA} simulations have reported that the mass of metals within stars in haloes over the total mass of metals in the same halo are respectively $\approx60\%$ and $\approx70\%$ \citep{Appleby2021, Mitchell2022b}. If we do the same calculation, we find that our simulations bracket theirs ($\approx80\%$ for \KI\ and \KR, $\approx50\%$ for \DC). However, it is more instructive to look at the mass of metals in the different regions of the halo over the mass of metals \textit{released} by stars in the halo than only the metals within the halo. 
        We show this in the top panel of Fig.~\ref{fig:PIES_metal}, with the location of all the metals produced by the galaxy. More specifically, we count the mass of metals locked up in stars $M_\mathrm{Z,star}$, the mass of metals in the ISM $M_\mathrm{Z,ISM}$ (i.e. with $r < 0.1 R_\mathrm{200, min}$) and within the CGM $M_\mathrm{Z,CGM}$ (i.e. with $0.1 R_\mathrm{200, min} < r < R_\mathrm{200, min}$). We normalise these quantities to the total mass of metals produced by the stars in the simulated galaxy $M_\mathrm{Z,prod}$, and compute the mass of metals that have been ejected out of the DM halo as $M_\mathrm{Z,out} = M_\mathrm{Z,prod} - M_\mathrm{Z,star} - M_\mathrm{Z,ISM} - M_\mathrm{Z,CGM}$. We refer to them as \textit{missing metals}. On average, the mass of metals produced by the galaxy over the redshift range z=1.3-1 is $M_\mathrm{Z,prod,\KI}=6.7\times10^8\rm\ M_\odot$ for \KI, $M_\mathrm{Z,prod,\KR}=7.6\times10^8\rm\ M_\odot$ for \KR\ and $M_\mathrm{Z,prod,\DC}=3.2\times10^8\rm\ M_\odot$ for \DC.
        Strikingly, the galaxies in \KI\ and \KR\ retain $\approx90\%$ of the metals they produce within their stars or ISM, while the \DC\ galaxy only keeps $\approx25\%$ of its metal production within $0.1R_\mathrm{200, min}$. 
        The very large fraction of ejected metals in the \DC\ simulation makes it clear that the difference in the fraction of metals in stars or the ISM is due to \DC\ very efficiently expelling metals from their production sites to large distances, beyond the virial radius, from where they do not return. This is possibly thanks to a collective enhancement of the supernovae (the supernovae in a stellar particle all occur at the same time in \DC) and to the model producing stronger and hotter outflows \citep{Rosdahl2017}. \KI\ and \KR\ exhibit comparable behaviour regarding metal distribution. With \KI's feedback, metals that leave the ISM also mostly leave the DM halo for good, such that there is only a small fraction of metals present in the CGM. In \KR, there is only a small fraction of metals in the CGM, but the feedback also ejects very little matter beyond the virial radius. Despite these differences, both \KI\ and \KR\ exhibit a similar total amount of metals in the CGM ($2.3\% \times {M_\mathrm{Z,prod,\KI}}=1.54\times10^7\rm\ M_\odot$ and $2.1\% \times {M_\mathrm{Z,prod,\KR}}=1.59\times10^7\rm\ M_\odot$). The CGM in \DC\ exhibits both a high gas content and a high metallicity, resulting in a higher total amount of metal ($12.5\% \times M_\mathrm{Z,prod,\DC}=3.95\times10^7\rm\ M_\odot$). Despite a disparate fraction of baryons in the CGM, these results are within the scatter found in the EAGLE simulations. Our results are however inconsistent with IllustrisTNG, in which the CGM gas mass fraction is predominantly higher than 30\% for similar halo masses \citep{Davies2020}.

        \begin{figure}
            \begin{subfigure}[h]{\columnwidth}
        	       \includegraphics[width=\columnwidth]{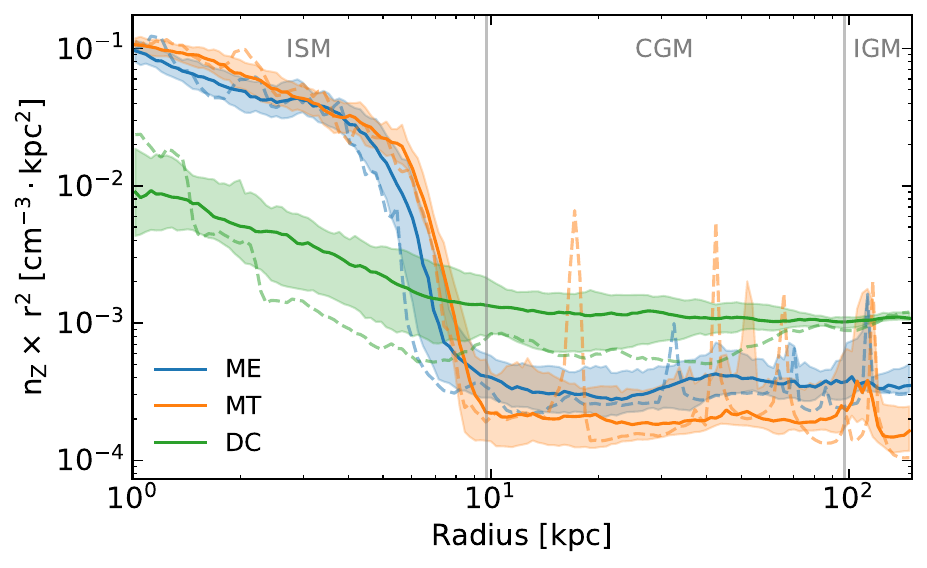}
            \end{subfigure}
            \caption{Mass of metals as a function of radius for each simulation. We bin the data radially and plot the median over $1\rm\ Gyr$ (from $z=1.3$ to $z=1$) in solid lines. The shaded areas correspond to the 15.9 and 84.1 percentiles. We also show with dashed lines the radial mass of metals in the last snapshot, at $z=1$. Lastly, the figure is split between the ISM, the CGM and the IGM with two grey lines placed at $0.1\ R_{200}$ and $R_{200}$.}
            \label{fig:radial_Z}
        \end{figure}

        Observers often characterise the CGM through radial column densities, and all ions commonly used for CGM measurements depend on metallicity. Understanding the distribution of metals is thus crucial in understanding the CGM. Furthermore, we found in Fig.~\ref{fig:PIES_metal} that while the distributions are different, the metal mass also varies a lot between the three models due to different feedback modes, with \DC\ having much more metals in the CGM than \KI\ and \KR. In Fig.~\ref{fig:radial_Z}, we finally look at the radial density distribution of metals in the CGM. We find that the metal mass in \DC\ is much lower than in \KI\ and \KR\ in the galaxy, yet stands much higher than \KI\ and \KR\ in the CGM. Similarly to the total radial gas density observed in Fig.~\ref{fig:radial_nH}, we observe that the radial distribution of metals is more continuous in \DC, while we notice a large drop at the galaxy edge in \KI\ and \KR. The absence of such a decline in \DC\ shows how the galaxy in \DC\ is unable to retain its metals in the ISM against the powerful feedback produced, unlike in \KI\ and \KR. Although \KI\ and \KR\ have a similar fraction of metals in the CGM (respectively 2.3\% and 2.1\%, cf Fig.~\ref{fig:PIES_metal}), \KI\ exhibits higher metal densities in the CGM than \KR. This is due to both the spherical binning and the temporal evolution of the CGM. The solid lines correspond to the binned median over all timesteps within $1.3>z>1$. We also show in dashed lines the metal distribution of the last output and find that metals are indeed especially concentrated around satellites in \KR, while \KI\ has an overall higher metal density in the CGM, yet lower metal densities around its satellites. As the satellites fall into the central galaxy, the metal distribution is smoothed in \KR\ and brought to similar values as in \KI. In the last snapshot, all three simulations are on the lower part of the shaded areas due to recent lower star formation rates and a thus lower supply of metals. The lines at higher redshift, e.g. at $z=1.3$ (the first snapshot of our stack) are, conversely, higher than the shaded area in all three simulations.

    \subsection{Different feedback modes}
        Feedback can change the galaxy and its CGM properties in multiple ways, depending on how outflows affect the gas they interact with \citep{Faucher-Giguere2023}. We mainly distinguish two feedback modes in the form of preventive and ejective feedback \citep{Christensen2016, Mitchell2022}. While ejective feedback removes fuel for star formation by expelling gas out of the galaxy, preventive feedback impedes gas infall before it can reach the galaxy. 
        With an extreme case of preventive feedback, one could expect to find a metal-rich, high-pressure CGM filled with ejected gas, keeping pristine gas at larger distances and shutting down accretion. With extreme ejective feedback, one could expect the opposite behaviour, with all the enriched gas ejected out of the halo while pristine gas can freely infall. In reality, it is likely that both regimes coexist in a complex manner, and the final state depends on how the gas phases will mix, and on the scales over which this feedback occurs. For example, if preventive feedback were effective at the galactic scale, one could expect to find a significant amount of accumulated pristine gas in the CGM. On the other hand, if feedback prevented gas infall further than the halo scale, the CGM could effectively be devoid of gas. 
        With these arguments in mind, it is possible to discuss feedback modes through which these models affect the properties of the galaxies and their CGM.

        Simulations \KI\ and \DC\ have a higher fraction of missing metals than that of missing baryons in their haloes. This suggests that the \textit{ejection} of matter outside of the haloes by galactic winds is the driving process that regulates the baryon fraction in these simulated haloes. On the contrary, \KR\ shows a relatively large missing baryon fraction ($\approx20\%$) but seems to retain most of its metals within $R_{200}$ ($\approx94\%$). This suggests that outflows are not efficient at pushing gas outside of the halo in this simulation, but somehow manage to slow down accretion of pristine material (see Figs.~\ref{fig:radial_outflows} and \ref{fig:radial_inflows}). 
        There is an additional complexity due to the fraction of metals locked up in stars. \KI\ and \KR\ both retain a large fraction of metals in stars. This suggests that metals released by stars are retained long enough in the ISM to form new stars because supernovae feedback is not strong enough to lower densities in the central region, allowing metals to mix in the ISM. On the contrary, \DC\ keeps only a small fraction of its metals in its stars due to strong feedback that ejects a large amount of metals from the galaxy. This reminds us that accurately modelling outflows in simulations is difficult not only because of the numerically demanding nature of physical processes that may happen in the CGM \citep{Faucher-Giguere2023}, but also because details of metal mixing at the launch sites of galactic winds determine the metallicity profile of outflows \citep{MacLow1999}.

        As \KI\ and \DC\ share the same star formation models, the difference in metal ejection is solely driven by the feedback model. Delayed cooling is known to produce stronger outflows \citep{Rosdahl2017}, but it also forbids immediate star formation by keeping the surrounding media at high effective pressure. This leads to burstier star formation and an overall more powerful feedback than in both other models. This feedback is hence more efficient in driving far-reaching outflows, which directly correlates with a higher efficiency in driving metals out of the halo.
        While the star formation and feedback models in \KI\ and \KR are relatively similar, they nonetheless produce distinct feedback modes. By resimulating a star formation peak in similar simulations, \citet{Rey2022} showed that they form stars in contrasting ways, either driven by turbulence or gravitation. Ensuing supernovae will thus explode in disparate media and interact with the ISM gas differently. Also, while \KI\ only injects feedback energy as momentum in both resolved and unresolved regimes, \KR\ also injects thermal energy in all cases, assuming it will dissipate due to the lack of resolution if momentum is also injected. With a different threshold to distinguish the two regimes and even different equations for the momentum injected, these additional distinct features in both models can collectively drive contrasting feedback modes. The most appropriate approach to conclusively determine how the feedback modes are driven would be to perform simulations of the three models in an idealised environment; however, this analysis lies beyond the scope of the present study.

\section{Conclusions} \label{sec:conclusions}
In this work, we run cosmological zoom-in simulations of the same galaxy with three different subgrid models for star formation and feedback using the radiation hydrodynamical code \textsc{RAMSES-RT}. We refer to our three models as \KI\ \citep{Kimm2017}, \KR\ \citep{Kretschmer2020} and a variation of \KI\ with supernova feedback modelled through delayed cooling \citep{Teyssier2013}, \DC. We have seen that three simulations with different feedback models can be calibrated to produce galaxies with similar stellar masses. Despite these similar stellar properties, we find that feedback from supernovae operates very differently in each model, and leads to varying CGM properties. Our results can be summarised as follows.

\begin{itemize}
    \item[$\bullet$] \textbf{Three different star formation and stellar feedback subgrid models can be calibrated in stellar mass.}
        The three models \KI, \KR, and \DC, often used in the literature, are employed on the same set of initial conditions and are calibrated to produce a similar stellar mass within a factor of approximately two. 
    \item[$\bullet$] \textbf{The burstiness is sensitive to subgrid models.}
        Thanks to a more powerful feedback, \DC\ exhibits much higher burstiness than both \KI\ and \KR\ at all timescales considered. \KI\ exhibits similar burstiness on both short and long timescales, while \KR\ is burstier than \KI\ and shows a higher burstiness when computing it on larger timescales, possibly due to the implementation of subgrid turbulence in \KR and the different properties of the star-forming media.
    \item[$\bullet$] \textbf{The halo gas of the three simulations shows similar properties, but disparate distributions.}
        The phase diagram of the three simulations is comparable with three phases: the cold dense ISM phase, the hot and low-density ejected or shock-heated phase, and the warm phase plateauing at $T\sim10^4\rm\ K$. However, while \KI\ and \KR\ have a similar amount of gas in the ISM and in the CGM, \DC\ has much more gas in the CGM, and only a small fraction of cold gas in the ISM as of its powerful feedback strips most of the gas from the ISM and the model itself heats a fraction of the remaining gas by construction.
    \item[$\bullet$] \textbf{Inflows and outflows are dominated by hot gas in all simulations, but warm gas shows contrasting behaviours.}
        In all three simulations, both inflows and outflows are dominated by hot gas which entrains gas from the CGM and reduces the inflow rate of hot gas. However, in the inner CGM, the total inflow rate is dominated by warm gas in \DC. The flows in \DC\ exhibit signs of warm gas recycling, with both inflows and outflows decreasing as a function of radius. In \KI\ and \KR, warm outflows are sporadic in time but have otherwise comparable rates to \DC.
    \item[$\bullet$] \textbf{Feedback leads to drastically different baryon distributions in the halo.}
        Through feedback much more efficient in removing gas from the halo, the baryonic content of the halo in \DC\ is reduced $\approx56$\% of the universal fraction, and to $\approx79$\% and $\approx94$\% in \KR\ and \KI. The distribution of this gas is also different among the three simulations, \DC's ISM and stellar mass each constitute less than half the fractions found with \KI\ and \KR. The CGM gas fractions are also distinct, with \DC\ exhibiting the highest fraction and \KR\ the lowest.
    \item[$\bullet$] \textbf{Feedback leads to drastically different metal distributions in the halo.}
        The metal mass in the ISM is much lower in \DC\ than in \KI\ and \KR, while it is much higher in the CGM. \DC\ feedback manages to eject 65\% of the metal mass out of the halo while \KI\ and \KR\ only eject 5-10\%. Nonetheless, the galaxy's metallicity is compatible with observations in both \DC\ and \KI. Also, while \KI\ and \KR\ exhibit a similar total metal mass in the CGM, the radial mass of metals is higher in \KI\ as metals in \KR\ are concentrated around satellites.
    \item[$\bullet$] \textbf{Diverse feedback modes dominate the three simulations.}
        \DC\ is largely dominated by ejective feedback, leading to a metal and gas-rich CGM and a high fraction of metals (64\%) and baryons (44\%) ejected or kept out of the halo. Conversely, \KI\ and \KR\ are dominated by preventive feedback, keeping most metals in stars (72-75\%) and leading to a metal-poor CGM (2.3-2.1\%). \KR\ keeps pristine gas beyond the halo, reaching $\approx80\%$ of the universal baryon fraction inside the halo. \KI\ keeps pristine gas beyond the galaxy and thus exhibits a CGM filled with more gas, and baryon fraction close to the universal value. 
\end{itemize}

We have seen in this paper that feedback modes can regulate galaxy growth through distinct means, and produce galaxies with similar stellar masses. Nonetheless, we also found that even when producing comparable galaxies, the density, metallicity, and temperature of the CGM can greatly differ. Understanding the CGM is thus crucial to trace back and understand what dictates galaxy growth and the role of different feedback processes. As most simulations rely on distinct feedback models, it is timely and necessary to find new ways to distinguish subgrid models through new constraints extending beyond the properties of the galaxies. One such constraint exists in the form of quasar absorption lines. In this series of papers, we will investigate how our state-of-the-art simulations compare against these constraints, and examine whether CGM absorption lines can help distinguish subgrid models in the near future.

\section*{Acknowledgements}
We thank the anonymous referee for their suggestions. We gratefully acknowledge support from the CBPsmn (PSMN, Pôle Scientifique de Modélisation Numérique) of the ENS de Lyon for the computing resources. The platform operates the SIDUS solution \citep{Quemener2013} developed by Emmanuel Quemener. MR and TK were supported by the National Research Foundation of Korea (NRF) grant funded by the Korean government (RS-2022-NR070872 and RS-2025-00516961). TK is also supported by the Yonsei Fellowship, funded by Lee Youn Jae.
The supercomputing time for part of the numerical tests was kindly provided by KISTI (KSC-2024-CRE-0200), and large data transfer was supported by KREONET, which is managed and operated by KISTI.
This research has made use of the Astrophysics Data System, funded by NASA under Cooperative Agreement 80NSSC21M00561. We also acknowledge the use of Python \citep{VanRossum1995}, Matplotlib \citep{Hunter2007}, NumPy \citep{Harris2020} and Astropy \citep{AstropyCollaboration2013, AstropyCollaboration2018} for this work.


\section*{Data Availability}
The data generated and used in this article will be shared upon reasonable request to the corresponding author.


\bibliographystyle{mnras/mnras}   
\bibliography{ARCHITECTS_I} 


\bsp	
\label{lastpage}
\end{document}